\documentclass[11pt]{article}
\pdfoutput=1
\usepackage[utf8]{inputenc}
\usepackage{jheppub}

\usepackage{graphicx}
 \usepackage{subcaption}
\usepackage{amsmath, amsthm, amssymb}

\usepackage{comment} 

\usepackage{epsfig,latexsym,cancel}
\usepackage{amssymb,amsmath}
\usepackage{epstopdf}
\usepackage{hyperref, graphicx, slashed, subcaption}
\usepackage{multirow}
\usepackage[usenames, dvipsnames]{xcolor}
\usepackage{array}
\usepackage{booktabs}

\usepackage{epsfig,latexsym,cancel}
\usepackage{amssymb,amsmath}
\usepackage{epstopdf}
\usepackage{multirow}
\usepackage{array}
\usepackage{booktabs}
\usepackage{subcaption}

\newcommand{\GeV}{\ensuremath{\mathrm{GeV}}}
\newcommand{\brr}[1]{\left(#1\right)}

\title{A Stepping Stone Solution to the QCD Axion Isocurvature Problem}

\author{Kun-Feng Lyu,}
\emailAdd{kunfeng.lyu@ou.edu}
\author{Kuver Sinha}
\emailAdd{kuver.sinha@ou.edu}
\affiliation{Homer L. Dodge Department of Physics and Astronomy, University of Oklahoma, Norman, OK 73019, USA}

\abstract{
We propose a ``stepping stone'' realization of the varying Peccei Quinn (PQ) scale solution to the QCD axion isocurvature problem. In conventional implementations, suppressing isocurvature during high scale inflation requires a large inflationary PQ scale, while the subsequent evolution toward the much smaller late time vacuum can induce parametric resonance and nonthermal restoration of the PQ symmetry. We ameliorate this tension by introducing an intermediate PQ-breaking vacuum that is reached before the end of inflation, so that the later evolution to the present day PQ scale spans only a moderate hierarchy. The vacuum structure arises radiatively: a Yukawa coupling between the PQ scalar and PQ-charged fermions drives the quartic coupling negative at high scales while keeping it positive at lower scales. A positive dimension six operator then stabilizes a large field minimum, while the positive low energy quartic together with the quadratic terms generates the intermediate vacuum. We study two mechanisms in which the rolling inflaton removes the large field minimum, either by scanning a hidden non-Abelian gauge coupling or by directly scanning the Yukawa coupling. In both cases the PQ field is released toward the intermediate vacuum during inflation, after CMB modes have exited the horizon. We exhibit benchmark regions with subdominant backreaction on the inflationary sector and suppressed vacuum tunneling, and show that the construction permits substantially less hierarchical quartic and higher-dimensional couplings than conventional realizations that depend on a varying PQ scale. The underlying logic may have analogues for extra-dimensional or string axions, where the effective axion decay constant is controlled by geometric moduli, although realizing the required intermediate vacuum in a controlled modulus potential is nontrivial.
}

\begin{document}

\maketitle
\flushbottom

\section{Introduction}
\label{sec:introduction}

The QCD axion provides a simple and elegant solution to the strong CP problem~\cite{Peccei:1977hh,Peccei:1977ur,Weinberg:1977ma,Wilczek:1977pj,Shifman:1979if,Kim:1979if,Dine:1981rt,Zhitnitsky:1980tq}. It can emerge as the pseudo-Nambu Goldstone boson associated with the spontaneous breaking of the Peccei Quinn (PQ) symmetry at the scale $f_a$. Nonperturbative QCD effects generate an axion potential around the QCD confinement scale, and the axion relic abundance can be produced through the vacuum misalignment mechanism~\cite{Preskill:1982cy,Abbott:1982af,Dine:1982ah}.
Depending on the thermal history of the Universe, QCD axion cosmology can be classified into the pre-inflationary and post-inflationary scenarios~\cite{Marsh:2015xka,Fox:2004kb,DiLuzio:2020wdo}. In the pre-inflationary scenario, the PQ symmetry is broken during inflation and remains broken throughout reheating. The exponential expansion then stretches the axion field over our observable Universe, rendering its initial value nearly spatially homogeneous. By contrast, in the post-inflationary scenario, the reheating temperature is sufficiently high to restore the PQ symmetry, which is subsequently broken again as the Universe cools. Axion strings are then formed through the Kibble--Zurek mechanism~\cite{Kibble:1976sj,Kibble:1980mv,Zurek:1985qw}. Around the QCD confinement scale, the emergence of the axion potential leads to the formation of domain walls bounded by these strings~\cite{Vilenkin:1982ks,Kawasaki:2018bzv}. For a domain wall number $N_{\rm DW}>1$, a stable string domain wall network is formed, giving rise to the well-known domain wall problem. In the pre-inflationary scenario, such topological defects are inflated away, and the domain wall problem is therefore absent even for $N_{\rm DW}>1$. In this work, we focus on the pre-inflationary scenario. 

Assuming an initial axion misalignment angle $\theta_i=\mathcal{O}(1)$, the present-day axion relic abundance can be approximately written as~\cite{ParticleDataGroup:2026mpi}
\begin{equation}
\frac{\Omega_a}{\Omega_c}
\simeq
\gamma_{\mathrm{dil}}\,F_{\mathrm{anh}}\,
\theta_i^2
\left(
\frac{f_a}{9\times 10^{11}\mathrm{GeV}}
\right)^{1.165},
\end{equation}
where $F_{\rm anh}$ accounts for anharmonic corrections when $\theta_i$ is close to $\pi$, while $\gamma_{\rm dil}$ parametrizes a possible dilution of the axion abundance due to late-time entropy production. 
In the minimal misalignment scenario, this relation suggests a natural PQ scale of order $f_a\sim 10^{12}\,\mathrm{GeV}$ if the QCD axion constitutes all of the dark matter.

The pre-inflationary scenario is, however, subject to stringent constraints from axion isocurvature perturbations. For a representative value $f_a\sim 10^{12}\,\GeV$ and an order-one initial misalignment angle, current CMB observations typically require the inflationary Hubble scale to satisfy $H_{\rm inf}\lesssim 10^7\,\GeV$ in the minimal scenario~\cite{Verner:2026cbn,Petretti:2026ayw}, as reviewed in the next section. This bound disfavors high-scale inflation, which is predicted in many well-motivated inflationary models and can potentially be tested through searches for primordial tensor modes. The current BICEP/Keck bound, $r<0.036$~\cite{BICEP:2021xfz}, corresponds to
\begin{equation}
H_{\rm inf}\lesssim 4.5\times 10^{13}\,\mathrm{GeV}.
\end{equation}
Future experiments, including the AliCPT~\cite{Li:2017drr,Ghosh:2022mje}, Simons Observatory~\cite{SimonsObservatory:2025wwn}, LiteBIRD~\cite{LiteBIRD:2022cnt}, and CMB-S4~\cite{CMB-S4:2016ple,CMB-S4:2020lpa,CMB-S4:2022ght}, are expected to probe the tensor-to-scalar ratio down to the level of $r\sim 10^{-3}$. Consequently, a future detection of primordial gravitational waves at an observable level would place the minimal pre-inflationary QCD axion dark matter scenario under severe tension.

There are two broad approaches to alleviating the tension between high-scale inflation and the axion isocurvature constraint. The first is to make the axion sufficiently heavy~\cite{Dvali:1995ce,Jeong:2013xta,Choi:2015zra,Co:2018phi,Chakraborty:2025lyp} during inflation, such that its quantum fluctuations are suppressed on superhorizon scales. This typically requires an additional source of explicit PQ breaking~\cite{Co:2023mhe,Dine:2004cq,Higaki:2014ooa,Dine:2014gba,Kawasaki:2015lea} or an enhancement of the axion potential during inflation. For example, an increased QCD confinement scale during inflation~\cite{Takahashi:2015waa,Kearney:2016vqw,Buen-Abad:2019uoc,Jeong:2022kdr,Dvali:2026ceb,Sfakianakis:2026rge,Freese:2026xax} can provide such a mechanism. 
The second approach is to raise the effective PQ-breaking scale during inflation relative to its late-time value. In this case, a fluctuation of the canonically normalized axion field corresponds to a smaller fluctuation of the axion misalignment angle, thereby suppressing the resulting isocurvature perturbation. After inflation, the PQ scalar subsequently evolves toward its present-day vacuum expectation value, characterized by the decay constant $f_a$. This possibility was first proposed by Linde~\cite{Linde:1991km} and has since been explored in a variety of realizations~\cite{Jeong:2013xta,Linde:1990yj,Kasuya:1996ns,Kasuya:1997td,Folkerts:2013tua,Kawasaki:2013iha,Chun:2014xva,Fairbairn:2014zta,Nakayama:2015pba,Harigaya:2015hha,Kobayashi:2016qld,Allali:2022yvx,Graham:2025iwx,Tadepalli:2026mdc}. For extra-dimensional axions, a similar mechanism may arise from the dynamics of the compactification volume modulus or through couplings between the inflaton and the radion~\cite{Conlon:2022pnx,Choi:2026tup}. Unlike the first class of solutions, this approach allows the axion to remain a light Goldstone mode during inflation, with the suppression of isocurvature originating from the enlarged field-space radius of the PQ scalar.

If one opts for a solution to the isocurvature problem that relies on a varying PQ scale, an immediate challenge arises~\cite{Kasuya:1996ns,Kasuya:1998td,Kawasaki:2013iha}. Assuming that the radial component $\rho$ of the PQ field  begins near the Planck
scale (at a location that we will refer to as $\rho_{\rm min}$) and rolls directly to the ordinary PQ vacuum after inflation (which we will call $v_{\rm PQ}$), it
can undergo oscillations with large amplitude. 
These oscillations may amplify
both radial and angular fluctuations through parametric or tachyonic resonance. 
If these fluctuations become sufficiently large, the complex PQ field can traverse the origin and explore the full angular field space. 
The PQ symmetry may then be nonthermally restored, regenerating axion strings and domain walls.
The early lattice simulation~\cite{Kawasaki:2013iha} shows that such PQ symmetry restoration is unavoidable if $\rho_{\rm min}/v_{\rm PQ} > 10^4$. 
A recent lattice simulation~\cite{Graham:2025iwx} gives a lower critical ratio for the moving PQ minimum case. 
Although sufficiently mild radial evolution~\cite{Harigaya:2015hha} can avoid dangerous resonance, a direct transition across the full hierarchy required by high-scale inflation remains difficult.

The purpose of this paper is to explore what we call a ``stepping stone" solution to this direct transition problem. We construct a model to ameliorate the hierarchy between the initial large value of the PQ vacuum $\rho_{\rm min}$ and its final desired much smaller value $v_{\rm PQ}$ by introducing an
intermediate PQ vacuum $\widetilde{\rho}_{\rm min}$ (the ``stepping stone"):
\begin{equation}
    \rho_{\rm min}
    \quad\longrightarrow\quad
    \widetilde{\rho}_{\rm min}
    \quad\longrightarrow\quad
    v_{\rm PQ}.
    \label{eq:intro_three_vacua}
\end{equation}
The PQ field is stabilized in the  initial vacuum $\rho_{\rm min}$  during the part of inflation that is relevant for the CMB, thereby suppressing the observable axion isocurvature
perturbation. Before inflation ends, the PQ field is transferred to the intermediate vacuum $\widetilde{\rho}_{\rm min}$. If one engineers $\widetilde{\rho}_{\rm min}$ to  only be a couple of  orders of magnitude above the final PQ scale $v_{\rm PQ}$, the subsequent evolution after inflation begins from a much smaller field value. This ensures that this post-inflationary evolution and the solution to the QCD isocurvature problem is protected from destabilization due to  parametric resonance.

The first essential new ingredient introduced in our paper is therefore the controlled engineering of the intermediate vacuum $\widetilde{\rho}_{\rm min}$. The roadmap is as follows. The initial vacuum $\rho_{\rm min}$  is generated by the competition between a positive dimension six operator in the Lagrangian and a quartic coupling that is negative at high scales. The quartic coupling runs from negative to positive values at lower energies, due to a Yukawa coupling of the PQ scalar $\Phi$ to fermions $\psi$. The fermions may be charged under a hidden non-Abelian gauge group with gauge coupling $g_D$, although this is not essential at this stage. The intermediate vacuum $\widetilde{\rho}_{\rm min}$ is thus dynamically generated by a competition between the (now positive) quartic term and a negative mass term in the Lagrangian.

The transition from $\rho_{\rm min}$ to the stepping stone at  $\widetilde{\rho}_{\rm min}$ has to be managed carefully, since it may resurrect the parametric resonance problem. We achieve this by making the radiatively generated vacuum structure depend on the inflaton, which is thus the second essential ingredient of our work. 
We explore two possible models that couple the inflaton to the PQ sector. In the first model, we require a hidden non-Abelian gauge group under which the fermions $\psi$ are charged, and couple the inflaton to the hidden gauge kinetic term. The slow evolution of the inflaton changes the hidden gauge coupling at the cutoff scale. Through the coupled renormalization group equations, this variation modifies the Yukawa and quartic trajectories and gradually deforms the PQ potential. For a suitable inflaton trajectory, the initial large field minimum disappears after the observable CMB modes have left the horizon. The PQ field is then released toward the intermediate vacuum while inflation is still ongoing. In the second model, we do not require a hidden non-Abelian gauge group. Here, we directly couple the inflaton to the PQ charged fermions $\psi$ through a dimension 5 operator. Similar to the first model, as the inflaton evolves one scans different boundary values of the Yukawa coupling, which modifies the PQ field potential and removes the large field vacuum position.

The final transition from $\widetilde{\rho}_{\rm min}$ to $v_{\rm PQ}$ follows standard methods. A nonminimal coupling to the Ricci scalar raises the lower-field vacuum during inflation and keeps it well above the size of de Sitter
fluctuations. After inflation, the curvature contribution decreases and
the intermediate vacuum $\widetilde{\rho}_{\rm min}$ moves toward the final PQ vacuum $v_{\rm PQ}$.

We present numerical RG trajectories demonstrating that a modest change in the hidden gauge coupling can shift and ultimately remove the initial large-field vacuum. 
We also provide benchmark points for which the associated change in the PQ vacuum energy remains a small fraction of the inflationary energy density.
Furthermore, we show that quantum tunneling can be highly suppressed before the large-field vacuum disappears. The PQ field therefore leaves $\rho_{\rm min}$ through classical rolling toward $\widetilde{\rho}_{\rm min}$ rather than through vacuum decay.

We note that our solution may also apply to extra-dimensional and string axions, which do not arise from the spontaneous breaking of a four dimensional global symmetry, but rather descend from higher dimensional gauge fields. Their effective decay constants are controlled by geometric moduli~\cite{Reece:2025thc, Conlon:2022pnx,Choi:2026tup, Chakraborty:2025lyp} rather than by  the radial mode of a four dimensional PQ field.  Extending the stepping stone mechanism to such axions would require engineering an analogous sequence of controlled modulus vacua, which is itself a nontrivial stabilization problem. We make some preliminary comments on this possibility in Section~\ref{sec:extra_dimensional_axions}.

The paper is organized as follows. In
Sec.~\ref{sec:isocurvature_review}, we review the axion isocurvature
problem, the varying-PQ-scale solution, and the parametric-resonance
constraint. In Sec.~\ref{sec:model}, we describe the distinct three vacuum structure in different stages. In
Sec.~\ref{sec:inflaton_controlled_evolution}, the inflaton controlled evolution is introduced and numerical benchmarks
are presented. While
Sec.~\ref{sec:transition_dynamics} discusses the transition dynamics and
remaining open issues. We conclude in Sec.~\ref{sec:conclusion}.

\section{Review: Axion Isocurvature and Varying PQ VEV Solution}
\label{sec:isocurvature_review}

In this Section, we review the standard solution to the axion isocurvature problem based on a large PQ scale during inflation.
 We discuss why the radial
evolution of the PQ field after inflation is potentially dangerous, and how introducing a stepping stone ameliorates the parametric resonance problem.

\subsection{Axion Isocurvature in the Pre-Inflationary Scenario}
\label{sec:axion_isocurvature}

To fix our notation, we express the complex PQ field $\Phi(x)$ as
\begin{equation}
    \Phi(x)
    =
    \frac{\rho(x)}{\sqrt{2}}
    e^{i\theta(x)}.
    \label{eq:PQ_polar_review}
\end{equation}
Here $\rho$ is the radial mode (which we will sometimes call the saxion, although we do not necessarily work within a supersymmetric context) and $\theta$ is the angular mode. Once the
PQ symmetry is spontaneously broken, the radial field acquires a
nonzero expectation value,
\begin{equation}
    \langle\rho\rangle=v_{\rm PQ}.
\end{equation}
The canonically normalized angular field is
$a(x)=v_{\rm PQ}\,\theta(x)$
and its anomalous coupling to QCD is given by 
\begin{equation}
    \mathcal{L}
    \supset
    \frac{g_s^2}{32\pi^2}
    N_{\rm DW}\theta\,
    G^a_{\mu\nu}\widetilde{G}^{a,\mu\nu}.
    \label{eq:axion_QCD_coupling_review}
\end{equation}
The physical axion decay constant and axion angle are therefore
\begin{equation}
    f_a
    \equiv
    \frac{v_{\rm PQ}}{N_{\rm DW}},
    \qquad
    \theta_a
    =
    \frac{a}{f_a}
    =
    N_{\rm DW}\theta.
    \label{eq:fa_definition_review}
\end{equation}

Our focus in this paper is on the pre-inflationary scenario, in which the PQ symmetry is
broken during inflation and is not subsequently restored. We also assume
that the axion is effectively massless during inflation, $m_{a,{\rm inf}}\ll H_{\rm inf}$.
%
The canonically normalized axion field then acquires fluctuations
\begin{equation}
    \delta a
    \simeq
    \frac{H_{\rm inf}}{2\pi}
 \ .    \label{eq:axion_fluctuation_review}
\end{equation}
And for a fixed PQ scale, one obtains the fluctuation of the physical axion angle to be 
\begin{equation}
    \delta\theta_a
    \simeq
    \frac{H_{\rm inf}}{2\pi f_a}.
    \label{eq:axion_angle_fluctuation_review}
\end{equation}
After the QCD potential becomes important, the axion begins to oscillate
about its CP-conserving minimum. In the small angle regime, the
misalignment abundance is proportional to the square of the initial
angle,
$\rho_a\propto\theta_i^2$.
The corresponding axion density fluctuation is
\begin{equation}
    S_a
    \equiv
    \frac{\delta\rho_a}{\rho_a}
    \simeq
    2\frac{\delta\theta_i}{\theta_i}
    \simeq
    \frac{H_{\rm inf}}{\pi f_a\theta_i}.
    \label{eq:axion_isocurvature_review}
\end{equation}
If the axion constitutes only a fraction of the dark matter, the cold
dark matter isocurvature power spectrum is
\begin{equation}
    \mathcal{P}_{S}
    \simeq
    \left(
    \frac{\Omega_a}{\Omega_{\rm DM}}
    \right)^2
    \left(
    \frac{H_{\rm inf}}{\pi f_a\theta_i}
    \right)^2.
    \label{eq:isocurvature_power_review}
\end{equation}
On the other hand, the isocurvature fraction at the CMB pivot scale is defined by
\begin{equation}
    \beta_{\rm iso}
    \equiv
    \frac{
    \mathcal{P}_{S}
    }{
    \mathcal{P}_{\zeta}
    +
    \mathcal{P}_{S}
    }
    \label{eq:beta_iso_review}
\end{equation}
and the observational bound
$\beta_{\rm iso}\lesssim \beta_{\rm iso}^{\rm max}\simeq 0.038$~\cite{Planck:2018vyg}
implies that 
\begin{equation}
    H_{\rm inf}
    \lesssim
    \pi f_a\theta_i
    \left(
    \frac{\Omega_{\rm DM}}{\Omega_a}
    \right)
    \left[
    \frac{
    \beta_{\rm iso}^{\rm max}
    }{
    1-\beta_{\rm iso}^{\rm max}
    }
    \mathcal{P}_{\zeta}
    \right]^{1/2}.
    \label{eq:general_isocurvature_bound}
\end{equation}
Using the value
$\mathcal{P}_{\zeta}
    \simeq
    2.1\times10^{-9}$,
one then obtains
\begin{equation}
    H_{\rm inf}
    \lesssim
    2.9\times10^7\GeV
    \left(
    \frac{f_a}{10^{12}\GeV}
    \right)
    \left(
    \frac{\theta_i}{1}
    \right)
    \left(
    \frac{\Omega_{\rm DM}}{\Omega_a}
    \right).
    \label{eq:isobound_H}
\end{equation}
For an axion that constitutes all of the dark matter with
$f_a\sim10^{12}\GeV$ and $\theta_i\sim\mathcal{O}(1)$, this is far below
the Hubble scale associated with potentially observable primordial tensor
modes:
\begin{equation}
    H_{\rm inf}
    \simeq
    2.5\times10^{13}\GeV
    \left(
    \frac{r}{0.01}
    \right)^{1/2}.
    \label{eq:Hinf_tensor_relation}
\end{equation}
The conventional pre-inflationary QCD axion is therefore in strong
tension with models of high scale  inflation unless the axion constitutes only a
small fraction of the dark matter or the initial misalignment angle is tuned.

A standard solution to this problem is to make the radial PQ
expectation value during inflation larger than its value at late times. Denoting the effective axion decay constant during the stage of inflation relevant for the CMB by
\begin{equation}
    f_I
    \equiv
    \frac{\rho_I}{N_{\rm DW}}\,\,
\end{equation}
 the angular fluctuation generated at horizon crossing is
\begin{equation}
    \delta\theta_a
    \simeq
    \frac{H_{\rm inf}}{2\pi f_I}.
    \label{eq:angle_fluctuation_fI}
\end{equation}
The axion abundance at late times in this situation remains controlled by $f_a$, while the
isocurvature perturbation is suppressed by $f_I$, resolving the tension. More explicitly, one can rewrite the bound in Eq.~\eqref{eq:isobound_H} in terms of the inflationary decay constant as
\begin{equation}
    f_I
    >
    3.6\times10^4\,
    \theta_i^{-1}
    H_{\rm inf}.
    \label{eq:iso_fI}
\end{equation}
This condition only needs to hold while the modes relevant for the CMB are crossing the horizon.

Model implementations of this scenario have been widely studied, with a commonly utilized realization introducing a nonminimal coupling $\xi$ between
the PQ scalar and the Ricci scalar:
\begin{equation}
    \mathcal{L}
    \supset
    |\partial_\mu\Phi|^2
    +
    \lambda f_a^2|\Phi|^2
    -
    \lambda|\Phi|^4
    +
    2\xi R|\Phi|^2.
    \label{eq:standard_Ricci_PQ}
\end{equation}
During approximately de Sitter inflation, $R\simeq12H_{\rm inf}^2$, 
so that one has 
\begin{equation}
    2\xi R|\Phi|^2
    \longrightarrow
    24\xi H_{\rm inf}^2|\Phi|^2.
\end{equation}
If this contribution dominates the negative quadratic term, the
inflationary decay constant is approximately
\begin{equation}
    f_I
    =
    \frac{H_{\rm inf}}{N_{\rm DW}}
    \sqrt{
    \frac{24\xi}{\lambda}
    }.
    \label{eq:standard_fI_Ricci}
\end{equation}
A sufficiently large $\xi/\lambda$ can therefore satisfy
Eq.~\eqref{eq:iso_fI}.

\subsection{Post-Inflationary Evolution and Parametric Resonance}
\label{sec:parametric_resonance_review}

Models that attempt to solve the axion isocurvature problem by introducing a varying PQ scale effectively shift the issue from controlling fluctuations of the axion during inflation to controlling the dynamics of the radial (saxion) field after inflation. It turns out that this introduces another classic problem.

If the initial position of the saxion after inflation is far from the desired final vacuum, it can acquire substantial kinetic energy and undergo oscillations with large amplitude. These oscillations can amplify both radial and angular fluctuations through parametric or tachyonic resonance. Once the radial variance becomes comparable to the background PQ scale, different spatial regions can approach the origin in field space. At the same time, the axion phase
can become randomized over its compact field space. The PQ symmetry may
thus be nonthermally restored, leading to the formation of axion strings
and, at the QCD transition, domain walls.

A lattice analysis of a quartic PQ potential~\cite{Kawasaki:2013iha,Kawasaki:2026jen}
found that avoiding
nonthermal restoration approximately requires 
\begin{equation}
    \frac{|\Phi_i|
    }{v_{\rm PQ}} < 10^4,
\label{eq:old_PR_hierarchy_bound}
\end{equation}
assuming the matter dominated era after inflation. Here $|\Phi_i|$ denotes the
field value from which the axion field begins to roll. This condition is difficult to reconcile with high scale inflation if the PQ field remains near the Planck scale until the end of inflation.
For example, taking
\begin{equation}
    \rho_I
    \sim
    0.1M_{\rm pl},
    \qquad
    v_{\rm PQ}
    \sim
    10^{12}\GeV,
\end{equation}
gives a hierarchy much larger than that required in Eq.~\eqref{eq:old_PR_hierarchy_bound}.

One way to avoid  PQ symmetry restoration in the early radial evolution is to include a dimension six operator or higher order~\cite{Harigaya:2015hha}. The PQ potential is written as
\begin{equation}\label{eq:V_dim6}
    V(\Phi)
    =
    -\mu^2|\Phi|^2
    +
    \lambda|\Phi|^4
    +
    \frac{c_6}{M_{\rm pl}^2}|\Phi|^6
    -
    24\xi H^2|\Phi|^2.
\end{equation}
When the saxion field value is very large, the potential is dominated by the sextic term. 
During the oscillation by the sextic term, parametric resonance is inefficient because the oscillation frequency of the saxion tracks the Hubble expansion rate, leaving insufficient time for substantial Floquet growth before the quartic term becomes dominant. 
Explicitly, for the general potential $\phi^p$, the EOS parameter during the oscillation era is given by
\begin{equation}
    w = \dfrac{p-2}{p+2} \ .
\end{equation}
Therefore the scalar potential energy evolves like $a^{-3(1+w)} = a^{-6p/(p+2)}$, yielding the scalar oscillating amplitude $\phi(t) \sim a^{-6/(p+2)}$. For a certain comoving mode $k$, the ratio of the physical momentum to the oscillation frequency is
\begin{equation}
  \dfrac{k/a}{\omega_\phi} = \dfrac{k}{a\, \sqrt{V''(\phi)}} \sim \dfrac{k}{a \, \phi^{\,p/2-1}} \sim k\, a^{\frac{2(p-4)}{2+p}}  \ .
\end{equation}
Self-resonance in a quartic potential with $p = 4$ is typically more efficient because the system is conformally invariant. Both the physical momentum $k/a$ and the oscillation frequency scale as $a^{-1}$. Therefore, a mode that enters a Floquet instability band can remain there for many oscillations and accumulate substantial exponential growth. 
For a sextic potential with $p = 6$, however, the oscillation frequency decreases faster, $\omega_\phi\propto a^{-3/2}$, so $k/(a\omega_\phi)\propto a^{1/2}$. The modes therefore drift across and eventually leave the instability bands, limiting the total amplification.
Therefore, only when the potential becomes dominated by the quartic term, can parametric resonance start to be efficient. The interval of efficient quartic resonance can thus be delayed.

The field value $\Phi_4$ at which the quartic and sextic terms become comparable
is determined by
\begin{equation}
    \lambda|\Phi|^4
    =
    c_6
    \frac{|\Phi|^6}{M_{\rm pl}^2}.
\end{equation}
This gives
\begin{equation}
    |\Phi_4|
    =
    M_{\rm pl}
    \left(
    \frac{\lambda}{c_6}
    \right)^{1/2}.
    \label{eq:cond_dim6_quartic}
\end{equation}
The effective initial amplitude relevant for the quartic resonance stage
is then $|\Phi_4|$, rather than the original Planck-scale field value. A
sufficient condition inferred from the conventional analysis is, from Eq.~\eqref{eq:old_PR_hierarchy_bound},
\begin{equation}
    |\Phi_4|
    <
    10^4v_{\rm PQ}.
    \label{eq:dimension_six_PR_condition}
\end{equation}
 The model in Eq.~\eqref{eq:V_dim6} thus solves the isocurvature problem, but at the cost of substantial fine-tuning. 
For a Planck-scale inflationary vev and
$v_{\rm PQ}\sim10^{12}\GeV$, for example, satisfying this condition can require very small parameters,
\begin{equation}\label{eq:c6_parameter_literature}
    \lambda
    \lesssim
    10^{-14},
    \qquad
    c_6
    \sim
    10^{-10},
\end{equation}
up to the precise normalization of the operators. The mass term $\mu$ is $\mu \sim \sqrt{\lambda} v_{\rm PQ} \sim 100$ TeV, much smaller than the inflationary scale.

\section{Model: The Stepping Stone Vacuum}\label{sec:model}

Having reviewed the standard mechanism for solving the axion isocurvature problem, we now turn to our main proposal: the introduction of the stepping stone vacuum $\widetilde{\rho}_{\rm min}$. Indeed, as stated in the Introduction, the evolution of the saxion in our scenario proceeds in the following way:
\begin{equation}
    \rho_{\rm min}
    \quad\longrightarrow\quad
    \widetilde{\rho}_{\rm min}
    \quad\longrightarrow\quad
    v_{\rm PQ}.
    \label{eq:two_stage_motivation}
\end{equation}
Here $\rho_{\rm min}$ denotes the initial large-field vacuum,
$\widetilde{\rho}_{\rm min}$ denotes the intermediate vacuum reached during
inflation, and $v_{\rm PQ}$ denotes the final  PQ-breaking scale.

The first transition occurs during inflation, after the CMB modes exit  the horizon. The second occurs after inflation. The three scales will be  chosen to satisfy
\begin{equation}
    \rho_{\rm min}
    \sim
    \mathcal{O}(0.1)M_{\rm pl},
    \qquad
    \widetilde{\rho}_{\rm min}
    \sim
    \mathcal{O}(10^3)v_{\rm PQ},
    \qquad
    v_{\rm PQ}
    =
    N_{\rm DW}f_a\,\,,
    \label{eq:three_scale_motivation}
\end{equation}
and they perform distinct roles:
\begin{align}
    \rho_{\rm min}
    &:\
    \text{suppresses the axion phase fluctuations generated on CMB scales},
    \nonumber\\
    \widetilde{\rho}_{\rm min}
    &:\
    \text{provides a resonance-safe field value at the end of inflation},
    \nonumber\\
    v_{\rm PQ}
    &:\
    \text{sets the late time axion decay constant and relic abundance}.
    \label{eq:three_scale_roles_review}
\end{align}
The first transition $ \rho_{\rm min}
    \longrightarrow
    \widetilde{\rho}_{\rm min}$ may span a large field hierarchy, but it occurs in
an inflating background, where Hubble friction and the remaining
inflationary expansion can damp the radial motion and dilute transition
products. The second transition takes place over a much smaller ratio which can satisfy the conventional constraints coming from parametric resonance.

The main feature that underlies our  construction of $\widetilde{\rho}_{\rm min}$ is the following: while the standard treatment reviewed in Section~\ref{sec:isocurvature_review} assumes that both the quartic coupling $\lambda$ and the Wilson coefficient $c_6$ of the dimension six operator are small and positive, in contrast we permit the quartic coupling $\lambda$ to be negative at high scales and positive at lower scales. In that case, the three vacua in Eq.~\eqref{eq:two_stage_motivation} have the following origins. The initial vacuum $\rho_{\rm min}$ is generated by the competition between a (\textit{negative}, at high scales)  quartic coupling and a positive dimension six operator. The intermediate vacuum $\widetilde{\rho}_{\rm min}$ is generated at lower
field values by the (now \textit{positive}, at lower scales) quartic term and the negative quadratic pieces, which includes the inflationary contribution from the coupling $\xi$ to the Ricci scalar. Finally, as the Hubble induced  contribution disappears after inflation, the intermediate vacuum $\widetilde{\rho}_{\rm min}$ continuously approaches the ordinary late time PQ
vacuum $v_{\rm PQ}$.

To engineer the required running of the quartic coupling $\lambda$ from negative to positive values, we consider a model where the PQ field $\Phi$ is coupled to a hidden sector fermion $\psi$. Although here we only require that one PQ fermion is charged under the strong group, one can introduce in general $N_F$ flavors of such fermions. The  potential for the PQ field is
\begin{equation}
\begin{split}
    V_{\rm PQ}(\Phi,H)
    ={}&
    -\mu^2|\Phi|^2
    -24\xi H^2|\Phi|^2
    +\lambda|\Phi|^4
    +\frac{c_6}{M_{\rm pl}^2}|\Phi|^6\,\,,
\end{split}
\label{eq:PQ_V_fermion}
\end{equation}
while the Yukawa interaction with the fermion $\psi$ is included through
\begin{equation}
    \mathcal{L}_{\rm Yuk}
    =
    -\left(
    y\,\overline{\psi}_L\Phi\psi_R
    +\mathrm{h.c.}
    \right).
    \label{eq:PQ_yukawa}
\end{equation}
Here $\psi$ is a PQ-charged fermion whose left and right handed
components carry opposite PQ charges. Although not essential at this stage, one may further assume that  $\psi$ transforms in the fundamental representation of a hidden non-Abelian gauge group $SU(N)$ with gauge coupling $g_D$:
\begin{equation}
    \mathcal{L}
    \supset
    -\frac{1}{4}
    \left(
    \frac{1}{g_{D}^2}
    \right)
    F^a_{\mu\nu}F^{a,\mu\nu}.
    \label{eq:1ferm_gauge_kinetic}
\end{equation}
The non-Abelian gauge sector in Eq.~\eqref{eq:1ferm_gauge_kinetic} will be essential for the model in Section~\ref{sec:inflaton_dependent_gauge_coupling}, while in Section~\ref{sec:inflaton_dependent_yukawa_coupling} it will be switched off. Here, we explore the general case where it exists.

As we shall see, the resulting RG evolution allows the quartic coupling to remain positive at lower scales while becoming negative at sufficiently high scales, reminiscent of the RG evolution of the Standard Model Higgs quartic, whose running
behavior is strongly affected by the top quark Yukawa.

The RG-improved potential of the saxion can be written as
\begin{equation}
    V_{\rm PQ}(\rho,H)
    =
    -\frac{1}{2}m_{\rm eff}^2(H)\rho^2
    +\frac{1}{4}\lambda(\rho)\rho^4
    +\frac{c_6}{8M_{\rm pl}^2}\rho^6 ,
    \label{eq:RG_improved_PQ_potential}
\end{equation}
where
\begin{equation}
    m_{\rm eff}^2(H)
    \equiv
    \mu^2+24\xi H^2.
    \label{eq:effective_PQ_mass}
\end{equation}
The parameters in Eq.~\eqref{eq:RG_improved_PQ_potential} are understood
as renormalized parameters. At one loop, the RG equations used in our
analysis are given by~\cite{Machacek:1983tz,Machacek:1983fi,Machacek:1984zw,Luo:2002ti}
\begin{align}
    \beta_{\lambda}
    \equiv
    \frac{d\lambda}{d\log\rho}
    &=
    \frac{1}{16\pi^2}
    \left(
    20\lambda^2
    + 4 N\lambda y^2
    -2 N y^4
    \right),
    \label{eq:beta_lambda}
    \\[0.5em]
    \beta_y
    \equiv
    \frac{dy}{d\log\rho}
    &=
    \frac{y}{16\pi^2}
    \left(
    (N+1)y^2- 3\dfrac{N^2-1}{N} g_D^2
    \right),
    \label{eq:beta_y}
    \\[0.5em]
    \beta_{g_D}
    \equiv
    \frac{dg_D}{d\log\rho}
    &=
    -\frac{g_D^3}{16\pi^2}
    \left(
    \frac{11N}{3}
    -\frac{2N_f}{3}
    \right).
    \label{eq:beta_gD}
\end{align}
Here $N_f$ is the number of active fermions charged under the hidden
$SU(N)$ group. In this work, we choose the benchmark value $N = 3$, $N_f = 12$ and the renormalization scale to be of order the
radial field value, $\mu_{\rm RG}\simeq\rho$, thereby resumming the
logarithms appearing in the large field effective potential. We neglect
the RG evolution of $\mu^2$ and $c_6$, since their variation is
subleading over the range of scales relevant to the benchmarks considered
below.

The boundary conditions are imposed at the cutoff scale $\Lambda$, which is assumed to be the Planck scale,
\begin{equation}
    \Lambda=M_{\rm pl}.
\end{equation}
The essential boundary condition distinguishing our construction from
the conventional positive-quartic scenario is
$\lambda(\Lambda)<0$.
The Yukawa contribution in Eq.~\eqref{eq:beta_lambda} permits
$\lambda(\rho)$ to be negative at high scales while becoming positive at
lower scales. The positive coefficient $c_6$ stabilizes the potential
where the quartic coupling is negative.
It will be useful to define the combination
\begin{equation}
    \overline{\lambda}(\rho)
    \equiv
    \lambda(\rho)
    +\frac{1}{4}\beta_\lambda(\rho).
    \label{eq:lambda_bar}
\end{equation}
This combination, rather than $\lambda(\rho)$ alone, enters the
stationary point condition for the RG-improved potential.

\subsection{Vacua Engineering}
\label{sec:large_vacuum}

In this Section, we perform calculations for the three vacua relevant for us.  We first consider the potential at field values close to the cutoff, aiming to engineer $\rho_{\rm min}$. In this regime, the quadratic term is parametrically small compared to
the quartic and dimension-six terms. Eq.~\eqref{eq:RG_improved_PQ_potential} therefore reduces to
\begin{equation}
    V_{\rm PQ}(\rho)
    \simeq
    \frac{1}{4}\lambda(\rho)\rho^4
    +\frac{c_6}{8M_{\rm pl}^2}\rho^6.
    \label{eq:large_field_PQ_potential}
\end{equation}
The negative quartic term tends to push the field toward large values,
while the positive dimension six term stabilizes the potential. Their
competition generates $\rho_{\rm min}$, which can be obtained as follows. Differentiating the  potential gives
\begin{align}
    \frac{dV_{\rm PQ}}{d\rho}
    \simeq
    \rho^3
    \left[
    \lambda(\rho)
    +\frac{1}{4}\beta_\lambda(\rho)
    +\frac{3c_6}{4M_{\rm pl}^2}\rho^2
    \right].
    \label{eq:large_field_derivative}
\end{align}
The position of the nonzero stationary point is therefore determined
implicitly by 
\begin{equation}
    \rho_{\rm min}
    =
    M_{\rm pl}
    \sqrt{
    -\frac{4\overline{\lambda}(\rho_{\rm min})}
    {3c_6}
    }.
    \label{eq:rho_min_cond}
\end{equation}
Its existence requires 
$\overline{\lambda}(\rho_{\rm min})<0$.
In addition, the stationary point must be a local minimum. Evaluating
the second derivative at $\rho=\rho_{\rm min}$ gives the condition
\begin{equation}
    -\overline{\lambda}(\rho_{\rm min}) + \dfrac{1}{2}\beta_\lambda(\rho_{\rm min})
    +\frac{1}{8}
    \left.
    \frac{d\beta_\lambda}{d\log\rho}
    \right|_{\rho=\rho_{\rm min}}
    >0.
    \label{eq:large_vacuum_stability}
\end{equation}

We now discuss $\widetilde{\rho}_{\rm min}$. At field values far below $\rho_{\rm min}$, the contribution from the dimension six operator  is
suppressed and the potential is controlled by the quadratic and quartic terms:
\begin{equation}
    V_{\rm PQ}(\rho,H)
    \simeq
    -\frac{1}{2}m_{\rm eff}^2(H)\rho^2
    +\frac{1}{4}\lambda(\rho)\rho^4.
    \label{eq:low_field_PQ_potential}
\end{equation}
We will choose the RG evolution such that the quartic coupling is positive in this field regime. The stationary point equation is
\begin{equation}
    \frac{dV_{\rm PQ}}{d\rho}
    =
    -m_{\rm eff}^2(H)\rho
    +\overline{\lambda}(\rho)\rho^3
    =0.
    \label{eq:low_field_stationary_condition}
\end{equation}
The nonzero minimum therefore satisfies
\begin{equation}
    \widetilde{\rho}_{\rm min}(H)
    =
    \frac{m_{\rm eff}(H)}
    {\sqrt{
    \overline{\lambda}
    \bigl(\widetilde{\rho}_{\rm min}\bigr)
    }},
    \label{eq:rho_tilde_min_general}
\end{equation}
where its existence requires
$\overline{\lambda}   \bigl(\widetilde{\rho}_{\rm min}\bigr)>0$.
Using Eq.~\eqref{eq:effective_PQ_mass} to substitute $m_{\rm eff}$ in Eq.~\eqref{eq:rho_tilde_min_general}, one obtains
\begin{equation}
    \widetilde{\rho}_{\rm min}
    \simeq
    \frac{
    \sqrt{\mu^2+24\xi H_{\rm inf}^2}
    }{
    \sqrt{
    \overline{\lambda}
    \bigl(\widetilde{\rho}_{\rm min}\bigr)
    }}.
    \label{eq:rho_tilde_min_inflation}
\end{equation}
This is the vacuum into which the PQ field is transferred after $\rho_{\rm min}$ disappears. We choose its position to be well above the characteristic inflationary
fluctuation,
\begin{equation}  \widetilde{\rho}_{\rm min}
    \gg
    \frac{H_{\rm inf}}{2\pi}.
    \label{eq:intermediate_scale_conditions}
\end{equation}

We finally turn to the final vacuum $v_{\rm PQ}$. The Ricci contribution decreases after inflation. In particular, for a
spatially flat FRW background dominated by a component with
equation-of-state parameter $w$,
\begin{equation}
    R=3H^2(1-3w)\,\,
    \label{eq:Ricci_FRW}
\end{equation}
which vanishes during radiation domination. The  potential at late times is
therefore obtained by setting the curvature-induced contribution to the effective mass to
zero. One obtains
\begin{equation}
    v_{\rm PQ}
    =
    \frac{\mu}{
    \sqrt{
    \overline{\lambda}(v_{\rm PQ})
    }}.
    \label{eq:final_PQ_vacuum}
\end{equation}
The physical  axion decay constant at late times is, correspondingly, 
$f_a = v_{\rm PQ}/N_{\rm DW}$.
Neglecting the mild running of the quartic coupling between the
intermediate and final vacuum scales, their ratio is approximately
\begin{equation}
    \frac{\widetilde{\rho}_{\rm min}}{v_{\rm PQ}}
    \simeq
    \sqrt{
    1+\frac{24\xi H_{\rm inf}^2}{\mu^2}
    }.
    \label{eq:intermediate_final_ratio}
\end{equation}
The parameters can therefore be chosen such that
\begin{equation}
    \frac{\widetilde{\rho}_{\rm min}}{v_{\rm PQ}}
    < 10^{4}.
    \label{eq:intermediate_final_benchmark_ratio}
\end{equation}
Unlike a direct transition from $\rho_{\rm min}$ to $v_{\rm PQ}$, the post-inflationary transition begins from the much lower scale
$\widetilde{\rho}_{\rm min}$.

The remaining problem is dynamical. The field must be transferred from $\rho_{\rm min}$ to $\widetilde{\rho}_{\rm min}$ after the modes relevant for the CMB  have exited the
horizon, but before inflation ends. In Section~\ref{sec:inflaton_controlled_evolution}, we will further augment our Lagrangian to include the inflaton-controlled operators. Two explicit operators will be discussed. The first one will  assume that the fermion is charged under a hidden non-abelian group, whose gauge coupling is controlled by the inflaton rolling. The second one will make the Yukawa coupling dependent on the inflaton field. 
These mechanisms will initiate a controlled evolution from $\rho_{\rm min}$ to $\widetilde{\rho}_{\rm min}$ 

\subsection{Inflationary Dilution and the Timing of the Transition}
\label{sec:transition_timing}

The transition $\rho_{\rm min} \rightarrow \tilde{\rho}_{\rm min}$ must occur after the CMB-relevant modes have exited
the horizon, so that their phase fluctuations are generated while
$\rho=\rho_{\rm min}$. 
Let $N_{\rm tr}$ denote the number of e-folds remaining after the
transition. The allowed ordering is
\begin{equation}
    0
    <
    N_{\rm tr}
    <
    N_{\rm CMB}.
    \label{eq:transition_timing_window}
\end{equation}
The lower bound is strengthened by the requirement of sufficient
inflationary damping.

For a coherently oscillating scalar in an approximately quartic
potential, the oscillation amplitude scales as
\begin{equation}
    \Delta\rho
    \propto
    a^{-1}.
    \label{eq:quartic_amplitude_damping}
\end{equation}
These estimates imply that even a substantial initial displacement can
be reduced exponentially in the number of e-folds:
\begin{equation}
    \frac{\Delta\rho_{\rm end}}
    {\Delta\rho_{\rm tr}}
    \sim
    e^{-pN_{\rm tr}},
    \qquad
    1\lesssim p\lesssim\frac{3}{2}.
    \label{eq:inflationary_radial_damping}
\end{equation}
For a hierarchy
\begin{equation}
    \frac{\rho_{\rm min}}
    {\widetilde{\rho}_{\rm min}}
    \sim
    10^2,
\end{equation}
one expects that a transition occurring at least 5 e-folds before the end of inflation may leave
sufficient time for substantial damping. The exact requirement depends
on the post-release potential and must be obtained from the numerical
trajectory.

\subsection{Avoiding PQ Symmetry Restoration} \label{sec:avoid_restore}

Since the vacua located at $\rho_{\rm min}$ and $\widetilde\rho_{\rm min}$ are occupied by the saxion during inflation, one has to ensure that quantum fluctuations of the saxion do not destabilize them.
The condition is
\begin{equation}
\langle \rho \rangle \gg \sigma_\rho
\equiv
\sqrt{\left\langle (\delta\rho)^2\right\rangle}\ .
\end{equation}
Here $\langle\rho\rangle$ denotes the position of the PQ minimum, which can be either $\rho_{\rm min}$ or $\widetilde{\rho}_{\rm min}$, while $\sigma_\rho$ is the root-mean-square fluctuation of the saxion field.

The fluctuation amplitude depends crucially on the effective saxion mass at the minimum. When $m_\rho = \sqrt{2}m_{\rm eff}\gtrsim H_{\rm inf}$, the saxion fluctuation is suppressed and decays after horizon exit. On the other hand, for $m_{\rho}\ll H_{\rm inf}$, the equilibrium variance in de Sitter space is approximately
\begin{equation}
\sigma_\rho^2
\simeq
\frac{3H_{\rm inf}^4}{8\pi^2m_{\rho}^2}\ .
\end{equation}
This expression assumes that the light-field stage lasts sufficiently long for the stochastic distribution to approach equilibrium.

For $\rho_{\rm min}$, $m_{\rho}\gg H_{\rm inf}$ is satisfied in the parameter space of interest. The quantum fluctuation around $\rho_{\rm min}$ is therefore strongly suppressed. At $\widetilde\rho_{\rm min}$, the saxion mass is approximately
$m_\rho = \sqrt{2} m_{\rm eff}(\widetilde{\rho}_{\rm min})
\simeq
4\sqrt{3\xi} H$.
If the saxion is light compared with the Hubble scale, its variance can be estimated as
\begin{equation}
\sigma_\rho
\simeq
\frac{H}{8\pi\sqrt{2\xi}}\,.
\end{equation}
Therefore, the condition
$\widetilde{\rho}_{\rm min}
\gg \sigma_\rho$
can be readily satisfied unless $\xi$ is extremely small. For a larger $\xi$, where $m_{\rho}(\widetilde{\rho}_{\rm min})$ becomes comparable to or larger than $H$, the fluctuation is further suppressed. We will confirm this behavior in the numerical analysis in Section~\ref{sec:inflaton_controlled_evolution}.

Another possible concern comes from fluctuation modes that become superhorizon during inflation. These modes may survive after inflation and later re-enter the horizon. If their amplitudes are sufficiently large, they may induce large spatial variations of the PQ field and lead to nonthermal PQ symmetry restoration.  The evolution of these modes is discussed in detail in the Appendix.
There are two possible sources of such fluctuations. The first is the quantum fluctuation generated during inflation. If
$m_{\rho}\ll H_{\rm inf}$ ,
the fluctuation generated around horizon crossing has the characteristic amplitude
\begin{equation}
\mathcal{P}_{\delta\rho}^{1/2}(k)
\sim
\frac{H_{\rm inf}}{2\pi}\ .
\end{equation}
For a sufficiently light saxion, the mode is almost frozen after horizon exit and can remain sizable until the effective mass becomes larger than the Hubble scale.

The second source is the parametric resonance associated with the rolling of the PQ field from $\rho_{\rm min}$ to $\widetilde{\rho}_{\rm min}$. During this stage, the oscillating background field can amplify fluctuations within certain momentum bands. This effect can be particularly efficient when the potential is dominated by the quartic term. 
If the amplified modes become superhorizon while the saxion remains light at $\widetilde{\rho}_{\rm min}$, their amplitudes can be preserved until the end of inflation.

These two types of fluctuations may enhance PQ symmetry restoration after inflation in two ways. Firstly, modes that exit the horizon shortly before the end of inflation can provide enhanced initial fluctuations for the subsequent rolling 
$\widetilde{\rho}_{\rm min}
\rightarrow
v_{\rm PQ}$.
The bound in Eq.~\eqref{eq:old_PR_hierarchy_bound}, which assumes vacuum-noise initial fluctuations, may then no longer be directly applicable. Even if the duration of the resonance is relatively short, the enhanced initial fluctuations may substantially reduce the time required for the system to become nonlinear.

Secondly, after the homogeneous PQ field reaches $v_{\rm PQ}$, there may still be a period during which
\begin{equation}
m_{\rho}(v_{\rm PQ})
=
\sqrt{2} m_{\rm eff}(v_{\rm PQ})
<
H(t) \ .
\end{equation}
During this period, the superhorizon fluctuations are not efficiently damped. When these modes re-enter the horizon, they can generate large local fluctuations around $v_{\rm PQ}$. If the fluctuation amplitude becomes $\sim v_{\rm PQ}$, the PQ symmetry is again restored nonthermally.

A conservative way to avoid this possibility is to require the saxion to remain heavier than the Hubble scale during inflation,
\begin{equation}\label{condpert2}
m_{\rho}(\rho_{\rm min})
\,\, \& \,\,  
m_{\rho}(\widetilde{\rho}_{\rm min})
\gtrsim
H_{\rm inf}\ .
\end{equation}
The condition  $m_{\rho}(\rho_{\rm min}) \gtrsim
H_{\rm inf}$ should automatically satisfy both conditions in Eq.~\eqref{condpert2}, since otherwise the PQ field cannot easily roll from $\rho_{\rm min}$ to $\tilde{\rho}_{\rm min}$. The fluctuations generated during inflation then decay outside the horizon instead of remaining frozen. We note that  $m_{\rho}(\rho_{\rm min}) \gtrsim
H_{\rm inf}$ is a sufficient but not necessary condition. A lighter saxion may still be allowed if the resulting fluctuation amplitude always remains much smaller than the instantaneous PQ field value, which needs to be confirmed by numerical simulations.


\subsection{Comments on Extra-Dimensional and String Axions}
\label{sec:extra_dimensional_axions}

The axion isocurvature problem is not restricted to axions arising from
the phase of a four dimensional PQ scalar. In extra-dimensional and
string constructions, axions may instead descend from higher dimensional
gauge fields, and their effective decay constants can depend on
geometric moduli. Schematically, for a dimensionless axion angle
$\theta_a$ one may write
\begin{equation}
    \mathcal{L}
    \supset
    -\frac{1}{2}K(\chi)(\partial\theta_a)^2,
\end{equation}
where $\chi$ denotes a modulus controlling the axion kinetic
normalization. A large value of the corresponding effective decay
constant during inflation can suppress axion isocurvature in much the
same kinematical way as a large radial PQ expectation value.

It is therefore natural to ask whether the stepping stone idea developed
in this work could have an analogue in such theories. At a schematic
level, one would seek a cosmological evolution in which the modulus
controlling the axion kinetic term passes through an intermediate
configuration before reaching its late-time value. Such a history could
separate the large inflationary axion scale required for isocurvature
suppression from the final relaxation to the present-day theory.

We emphasize, however, that our explicit construction does not establish
such a mechanism for extra-dimensional or string axions. The central
ingredient of the present model is the controlled engineering of several
vacua for the field that sets the axion scale. In a four dimensional PQ
theory this can be achieved through the RG-improved polynomial potential
studied above. For a geometric modulus $\chi$, by contrast, generating a
controlled potential is already a nontrivial problem. Obtaining an
additional metastable or intermediate minimum with the required
hierarchy, while maintaining control over the compactification and
avoiding destabilization of other moduli, introduces further constraints.

Moreover, the dynamical obstruction need not be the same as in the PQ
scalar realization. In the present model, violent evolution can restore
the PQ symmetry and regenerate topological defects. Extra-dimensional
axions need not possess an analogous four dimensional PQ restoration
transition. The relevant concerns are instead the excitation and
destabilization of the modulus sector, the evolution of the axion kinetic
matrix, and the preservation of the desired  compactification at late times. A
concrete realization of the stepping stone would require an explicit modulus stabilization
scenario in which the required intermediate vacuum can be generated and its cosmological evolution controlled.

\section{Inflaton-Controlled Evolution from $\rho_{\rm min}$ to $\widetilde\rho_{\rm min}$}
\label{sec:inflaton_controlled_evolution}

The existence of these two distinct vacua $\rho_{\rm min}$ and $\widetilde{\rho}_{\rm min}$ during inflation does not by itself determine the
cosmological evolution. If the initial vacuum remains stable throughout
inflation, the PQ field remains trapped at $\rho_{\rm min}$. If it is
metastable, the field may instead reach the lower-field vacuum through
quantum tunneling. Such a transition proceeds through bubble nucleation
and can generate large inhomogeneities during bubble expansion and
collision. In particular, fluctuations of the radial and angular
components of $\Phi$ may drive the field close to the origin, potentially
restoring the PQ symmetry and producing axion strings.

We instead seek a controlled classical transition in which the
large field vacuum $\rho_{\rm min}$ is removed by the evolving inflationary background. We will explore two models. In the first, discussed in Section~\ref{sec:inflaton_dependent_gauge_coupling}, our main idea is to introduce a hidden non-Abelian sector under which the fermions $\psi$ are charged, and require that the gauge fields have a dilatonic coupling to the inflaton $\phi$. In this setup, even though the inflaton does not couple directly to the PQ scalar, it changes the shape of the PQ scalar potential  in the following way: the UV boundary condition for the hidden
gauge coupling changes as the inflaton background evolves, which affects the coupled RG equations displayed in Eq.~\eqref{eq:beta_gD}, modifying the
Yukawa and quartic couplings. In the second model, discussed in Section~\ref{sec:inflaton_dependent_yukawa_coupling}, we instead make the Yukawa coupling depend on the inflaton. As we shall see, a  dynamical Yukawa coupling can alter the running quartic coupling, thereby changing the PQ field potential.

\subsection{Inflaton-Dependent Hidden Gauge Coupling}
\label{sec:inflaton_dependent_gauge_coupling}

We introduce a coupling between the inflaton $\phi$ and the kinetic term
of the hidden $SU(N)$ gauge sector:
\begin{equation}
    \mathcal{L}
    \supset
    -\frac{1}{4}
    \left(
    \frac{1}{g_{D,0}^2}
    +
    \kappa\frac{\phi}{\Lambda}
    \right)
    F^a_{\mu\nu}F^{a,\mu\nu}.
    \label{eq:inflaton_gauge_kinetic}
\end{equation}
Here $\Lambda$ denotes the cutoff scale, which will be taken to be of
order $M_{\rm pl}$ in the numerical analysis, and $\kappa$ is a
dimensionless coefficient. The parameter $g_{D,0}$ is the hidden gauge
coupling at the cutoff scale when $\phi=0$. 
The inflaton background acts as a dilaton-like field controlling the
normalization of the hidden gauge kinetic term. Such couplings 
can arise either from a renormalizable ultraviolet completion in four-dimensional spacetime or from an extra-dimensional scenario. For a comprehensive recent review of the origin of such couplings, we refer to ~\cite{deLima:2026xmt}.

After canonical normalization of the gauge field, the effective gauge coupling at the
cutoff scale is
\begin{equation}
    g_D(\Lambda,\phi)
    =
    \left(
    \frac{1}{g_{D,0}^2}
    +
    \kappa\frac{\phi}{\Lambda}
    \right)^{-1/2}.
    \label{eq:inflaton_dependent_gD}
\end{equation}
%
The positivity of the kinetic term requires
\begin{equation}
    \frac{1}{g_{D,0}^2}
    +
    \kappa\frac{\phi}{\Lambda}
    >0
    \label{eq:gauge_kinetic_positivity}
\end{equation}
throughout the inflationary trajectory. If $\kappa\phi$ decreases as inflation proceeds, the coefficient of
$F^a_{\mu\nu}F^{a,\mu\nu}$ decreases and the effective gauge coupling
increases. Conversely, an increasing value of $\kappa\phi$ weakens the
hidden gauge interaction. Its time variation is
\begin{equation}
    \frac{\dot g_D}{g_D}
    =
    -\frac{1}{2}
    g_D^2
    \frac{\kappa\dot\phi}{\Lambda}.
    \label{eq:gD_time_derivative}
\end{equation}
The sign and rate of the scan are therefore controlled by
$\kappa\dot\phi$.

We take the inflationary Hubble scale during the CMB-relevant window to
be
\begin{equation}
    H_{\rm inf}=10^{13}\GeV
    \label{eq:Hinf_benchmark}
\end{equation}
and assume an initial axion misalignment angle of order unity,
$\theta_i\sim\mathcal{O}(1)$. As stated before, we will choose $ \rho_{\rm min} \sim \mathcal{O}(0.1)M_{\rm pl}$.
%
%
%
The PQ sector must remain energetically subdominant so that its vacuum
evolution does not significantly disturb the inflationary background.
At the initial large-field minimum, the quadratic term is negligible,
and the PQ potential is approximately
\begin{equation}
    V_{\rm PQ}(\rho_{\rm min})
    =
    \frac{\lambda(\rho_{\rm min})}{4}\rho_{\rm min}^4
    +
    \frac{c_6}{8M_{\rm pl}^2}\rho_{\rm min}^6.
    \label{eq:PQ_energy_initial}
\end{equation}
We require
\begin{equation}
    \left|
    V_{\rm PQ}(\rho_{\rm min})
    \right|
    \ll
    V_{\rm inf}
    \equiv
    3M_{\rm pl}^2H_{\rm inf}^2.
    \label{eq:PQ_subdominance}
\end{equation}
For the representative value
$\rho_{\rm min}=0.2M_{\rm pl}$, 
this condition gives
\begin{equation}
    c_6
    \ll
    6.35\times10^{-6}
    -
    50\lambda(\rho_{\rm min}),
    \qquad
    \rho_{\rm min}=0.2M_{\rm pl}.
    \label{eq:c6_subdominance_bound}
\end{equation}
The dimensionless couplings are evolved according to
Eqs.~\eqref{eq:beta_lambda}--\eqref{eq:beta_gD}, with the boundary
conditions imposed at
$\Lambda=M_{\rm pl}$.
The (field-dependent) couplings entering the RG-improved potential are
obtained by integrating from $\Lambda$ to the renormalization scale
$\mu_{\rm RG}=\rho$.

We now turn to a discussion of the allowed range of values for the different parameters of our model. The parameters are the boundary values of the quartic and Yukawa couplings $\lambda(\Lambda)$ and $y(\Lambda)$, as well as the gauge coupling $g_D$ and parameter $c_6$. We will also comment on the level of fine-tuning required on the set $\{\lambda(\Lambda), y(\Lambda), g_D, c_6 \}$ compared to the values of $\lambda$ and $c_6$ required in Eq.~\eqref{eq:c6_parameter_literature}, which solves the isocurvature problem without introducing the stepping stone.

Firstly, the maximum value of the  quartic coupling is constrained by requiring that the potential energy contribution from the PQ sector remains sub-dominant to the inflationary potential energy. Defining $\Delta V_{\rm PQ}$ to be the difference in the PQ potential between the nearly disappearing large-field minimum and the intermediate minimum at $\widetilde{\rho}_{\rm min}$, we require 
\begin{equation}\label{vpqlessthanvinf}
    \frac{\Delta V_{\rm PQ}}{V_{\rm inf}}<0.2.
\end{equation}
This requirement is imposed to mitigate backreaction of the PQ sector on the inflationary sector and is discussed further in Section~\ref{backreactioninflaton}. 
 
 Under the requirement in Eq.~\eqref{vpqlessthanvinf}, we find that the largest allowed magnitude of the negative UV quartic coupling is approximately $|\lambda(\Lambda)|\simeq 7\times10^{-6}$. As far as the lower limit is concerned,  we have scanned  down to $\lambda(\Lambda)\simeq-2\times10^{-8}$ and checked that throughout this range, one can find  values of $c_6$ and $y(\Lambda)$  that realize the stepping stone mechanism. For $|\lambda(\Lambda)|\lesssim2\times10^{-8}$, however, the renormalization group evolution becomes increasingly sensitive to numerical precision. We therefore adopt $\lambda(\Lambda)=-2\times10^{-8}$ as the lower limit in our scans.

For a fixed $\lambda(\Lambda)$, the value of $c_6$ may be fixed by requiring a local minimum at $\rho=\rho_{\rm min}\simeq\mathcal{O}(0.1)M_{\rm pl}$. For fixed $\lambda(\Lambda)$, we find that there is a $\mathcal{O}(10\%)$ level of freedom in choosing $c_6$. Over the entire range of allowed $\lambda(\Lambda)$,  acceptable values of $c_6$ lie in the interval $[2.3\times10^{-7},1.2\times 10^{-4}]$, depending on the choice of the Yukawa coupling $y(\Lambda)$.  We scan the hidden gauge coupling over the range $g_D(\Lambda)\in[0.5,1]$. The required value of $y(\Lambda)$ is then determined by $\lambda(\Lambda)$ and $c_6$, and is found to lie in the range $[0.021,0.9]$. Thus, $y(\Lambda)$ is free to range with percent level variation.
This narrow range arises because $\lambda(\rho_{\rm min})$ is smaller than $\lambda(\Lambda)$ and is therefore highly sensitive to the $y^4$ contribution to its renormalization group evolution. If $y(\Lambda)$ is too small, increasing $g_D$ has little effect on the PQ vacuum structure, even when $g_D$ approaches unity. By contrast, a larger Yukawa coupling can eliminate the large field minimum directly.

Overall,  we thus obtain that the available range of the parameters is the following. The gauge coupling lies in the range $g_D(\Lambda)\in[0.5,1]$, while   $|\lambda(\Lambda)|$ ranges from $\sim \mathcal{O}(10^{-8})$ to around $7\times10^{-6}$. This leads $c_6$ and $y(\Lambda)$ to lie in the range  $[2.3\times10^{-7},1.2\times 10^{-4}]$ and $[0.021,0.9]$ respectively. We note that compared to the values required in  Eq.~\eqref{eq:c6_parameter_literature}, our mechanism does not require  an extremely small value of the quartic coupling $\lambda$; instead of $\lambda \sim 10^{-14}$, we can have values $\lambda \sim 10^{-7}$. Moreover, instead of requiring $c_6 \sim 10^{-10}$, one can have $c_6 \sim 10^{-4}$.

We now present our benchmarks. The first benchmark is chosen such that the PQ potential difference satisfies $\Delta V_{\rm PQ}/V_{\rm inf}\simeq\mathcal{O}(0.1)$, whereas the second benchmark is chosen to satisfy the smaller ratio $\Delta V_{\rm PQ}/V_{\rm inf}\simeq\mathcal{O}(0.01)$.

\subsubsection{Benchmark I}
\label{sec:initial_benchmark}

\begin{figure}[t]
    \centering
    \includegraphics[width=0.42\linewidth]{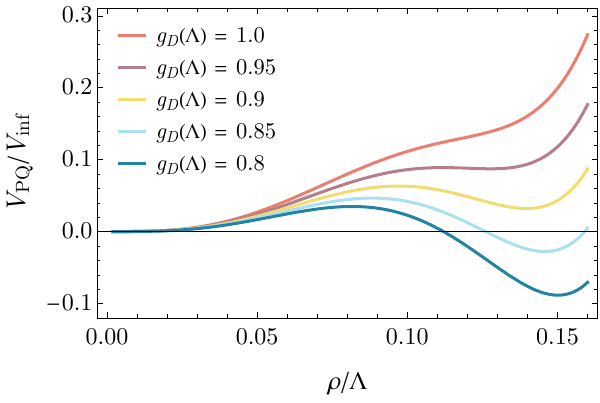}
    \includegraphics[width=0.415\linewidth]{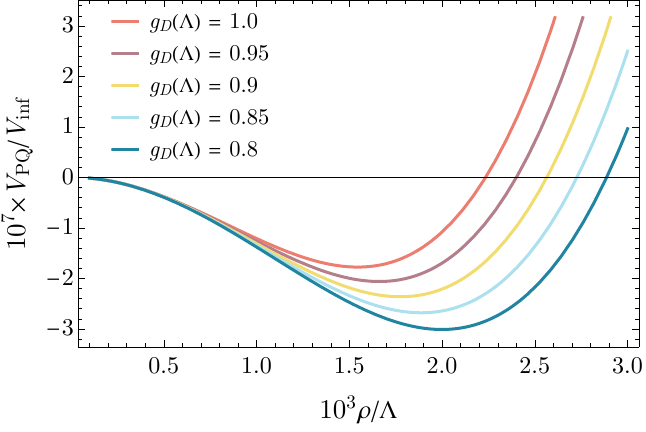}
    \caption{
    Left panel: The RG-improved PQ potential in the vicinity of the initial
    large field vacuum $\rho_{\rm min}$ for several boundary values of the hidden gauge
    coupling $g_D(\Lambda)$, for Benchmark I defined in Eq.~\eqref{eq:para_1}. The vertical axis is normalized to the
    inflationary energy density
    $V_{\rm inf}=3M_{\rm pl}^2H_{\rm inf}^2$. As
    $g_D(\Lambda)$ increases, $\rho_{\rm min}$ moves toward
    smaller values, becomes shallower, and eventually disappears. Right panel: The RG-improved potential in the vicinity of $\widetilde{\rho}_{\rm min}$ for the same set of values of $g_D(\Lambda)$.
    }
    \label{fig:V_gD_illustration}
\end{figure}

For the first benchmark, we choose
\begin{equation}
    c_6
    =
    4.5\times10^{-5},
    \qquad y(\Lambda) = 0.065, \qquad
    \lambda(\Lambda)
    =
    -2\times10^{-6}.
    \label{eq:para_1}
\end{equation}
while varying $g_D$.  For this benchmark, we choose parameters such that $V_{\rm PQ}/V_{\rm inf} \sim \mathcal{O}(0.1)$. Moreover, we choose $\xi = 1/48$ such that the saxion mass at $\rho = \widetilde{\rho}_{\rm min}$ is equal to $H_{\rm inf}$, in which the fluctuations can be safely diluted as discussed in Sec.~\ref{sec:avoid_restore}.  
The left panel of Fig.~\ref{fig:V_gD_illustration} shows the PQ potential in the vicinity
of the initial large-field minimum $\rho_{\rm min}$ for several values of the hidden
gauge coupling at the cutoff scale. The potential is normalized by the inflationary vacuum energy $V_{\rm inf}
    =
    3M_{\rm pl}^2H_{\rm inf}^2$.

The displayed field interval is restricted to the neighborhood of
$\rho_{\rm min}$ in order to focus on the evolution of the large-field
minimum.
As $g_D(\Lambda)$ increases, the position of the initial vacuum moves
toward smaller field values and its vacuum energy rises. 
The local minimum disappears when the boundary value
of the hidden gauge coupling becomes larger than approximately $g_D(\Lambda)\simeq 0.95$.
No solution to Eq.~\eqref{eq:rho_min_cond} exists for larger $g_D(\Lambda)$.  
The right panel of Fig.~\ref{fig:V_gD_illustration} displays the intermediate vacuum position $\widetilde{\rho}_{\rm min}$ for different $g_D(\Lambda)$. 
For $g_D(\Lambda) \simeq 1.0$, the  $\widetilde{\rho}_{\rm min}$ is around $1.6 \times 10^{-3} M_{\rm pl} = 3.9 \times 10^{15}\GeV$ which satisfies the PQ symmetry non-restoration condition of  Eq.~\eqref{eq:old_PR_hierarchy_bound}. 

\begin{figure}[t]
    \centering
    \includegraphics[width=0.99\linewidth]{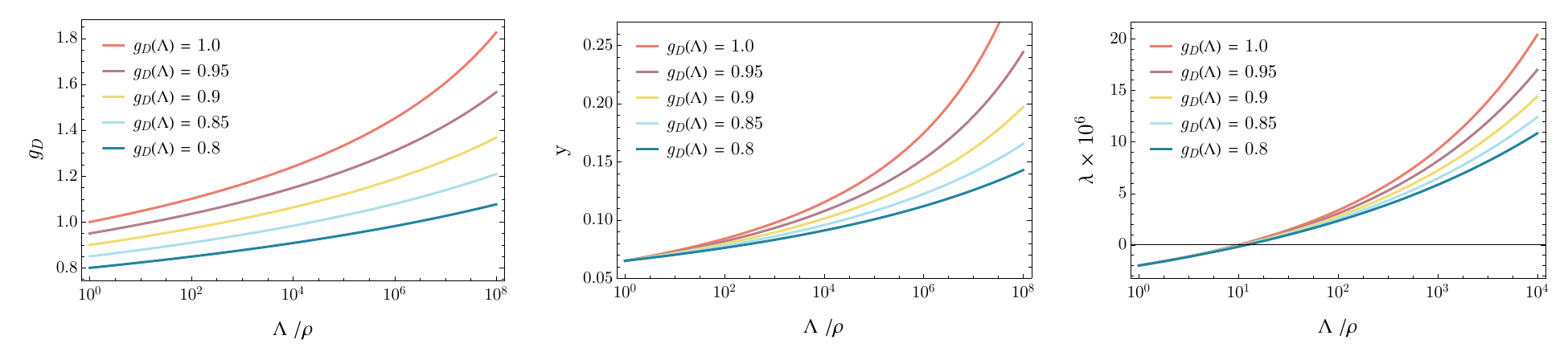}
    \caption{
    The running couplings $g_D(\rho)$, $y(\rho)$, and $\lambda(\rho)$
    as functions of $\Lambda/\rho$ for Benchmark I, defined in
    Eq.~\eqref{eq:para_1}. Moving to the right corresponds to running
    from the cutoff scale toward smaller field values. The different
    curves correspond to different boundary values of
    $g_D(\Lambda)$.
    }
    \label{fig:Couplings_illustration}
\end{figure}

To further understand the evolution of the potential, in
Fig.~\ref{fig:Couplings_illustration} we show  the running hidden gauge coupling,
Yukawa coupling, and the quartic coupling. The horizontal axis is chosen to
be $\Lambda/\rho$, so that moving to the right corresponds to evolving
from the cutoff scale $\Lambda$ toward smaller renormalization scales.
From the right panel, it is clear that 
the quartic coupling changes sign at a scale of order
\begin{equation}
    \rho_{\rm flip}
    \simeq
    0.1\Lambda
    \label{eq:rho_flip_first_benchmark}
\end{equation}
for our benchmark parameters. Increasing $g_D(\Lambda)$ shifts
$\rho_{\rm flip}$ toward larger field values. 
The region in which
$\lambda(\rho)$ is negative consequently contracts, and the condition
$\overline{\lambda}
\bigl(\rho_{\rm min}\bigr)<0$
needed to support the initial vacuum eventually fails.
In contrast, $\rho_{\rm min}$ gradually moves towards smaller values while $V_{\rm PQ}(\rho_{\rm min})$ increases. This behavior can be understood from the faster increase of $\lambda$. Near $\rho \simeq \rho_{\rm min}$, the combination $\bar{\lambda}=\lambda+\beta_\lambda/4$ is dominated by the beta-function contribution. A larger Yukawa coupling drives $\beta_\lambda$ more negative, requiring a larger positive $\lambda$ to compensate and thereby favoring a smaller $\rho_{\rm min}$.
Meanwhile, as can be seen from Eq.~\eqref{eq:PQ_energy_initial}, the larger value of $\lambda$ raises the PQ potential, resulting in a higher $V_{\rm PQ}(\widetilde{\rho}_{\rm min})$. 
Similarly, for the intermediate vacuum, 
$\widetilde{\rho}_{\rm min}$ also gradually moves towards smaller values and $V_{\rm PQ}(\rho_{\rm min})$ is lifted up, due to the faster increased $\lambda$ value. 

The results, as described till now, establish that the  boundary value
$g_D(\Lambda,\phi)$, which is controlled by the inflaton, can remove the initial large field minimum at $\rho_{\rm min}$. They do
not yet establish the precise trajectory followed by the PQ field after the minimum disappears, an issue  we will discuss in the subsequent Section~\ref{sec:transition_dynamics}. 
We can choose the following two representative boundary values of the hidden gauge coupling,
\begin{equation}
    g_{D,1}(\Lambda) = 0.9, \qquad g_{D,2}(\Lambda) = 1.0 \ ,
\end{equation}
in which $g_{D,1}(\Lambda)$  is the value at the inflationary CMB window and $g_{D,2}(\Lambda)$ corresponds to the value when the large field vacuum disappears at the subsequent inflationary era.
From Eq.~\eqref{eq:inflaton_dependent_gD}, one can obtain 
\begin{equation}
    \kappa
    \frac{\Delta\phi}{\Lambda}
    =
    \frac{1}{g_{D,2}^2(\Lambda)}
    -
    \frac{1}{g_{D,1}^2(\Lambda)}\,\,\,\, \simeq
    -0.23 \ .
\label{eq:gauge_scan_exact_relation1}
\end{equation}

Once the initial minimum disappears, the PQ field is released toward the
smaller values of $\rho$, and we eventually want to trap it at $\widetilde{\rho}_{\rm min}$, given by Eq.~\eqref{eq:rho_tilde_min_inflation}. 
We can perform an estimation on the possible parametric resonance coefficient based on~\cite{Kawasaki:2026jen}. 
Using the variable $z$ parameterizing the exponential growth rate, 
one obtains 
\begin{align}
    z
    &\equiv
    \frac{m_{\rm eff}}{H_{\rm inf}}
    \simeq
    \frac{\sqrt{\lambda}\,\rho_{\rm flip}}
    {H_{\rm inf}}
    \simeq
    \frac{
    10^{-3}\times10^{17}\GeV
    }{
    10^{13}\GeV
    }
    \left(
    \frac{10^{13}\GeV}{H_{\rm inf}}
    \right)
   =
    10
    \left(
    \frac{10^{13}\GeV}{H_{\rm inf}}
    \right)
    <100.
    \label{eq:z_first_transition}
\end{align}
Here we have used the fact that the quartic term quickly dominates the potential once the saxion field value decreases below $\rho_{\rm flip}$. Therefore we use the effective mass at $\rho_{\rm flip}$ as a good approximation.
This estimate suggests that the parametric resonance amplification is inefficient during the first transition and PQ symmetry is unlikely to be restored.  It should also be emphasized that the random seed amplitude for the sub-horizon fluctuation mode is suppressed thanks to $m_{\rm eff} \gg H_{\rm inf}$.

The estimate in Eq.~\eqref{eq:z_first_transition} should be
regarded as an approximate treatment for symmetry non-restoration.
The conventional condition $z<100$ was obtained assuming matter-dominated era after inflation. 
The present transition
occurs in an inflating background with nearly constant Hubble value and de Sitter quantum fluctuations.
A dedicated numerical simulation, which we leave for future work, is required to 
confirm that parametric resonance is indeed inefficient.

After inflation, the field approaches the final vacuum $v_{\rm PQ}$, given by Eq.~\eqref{eq:final_PQ_vacuum}. The parameters are chosen such that $\frac{\widetilde{\rho}_{\rm min}}{v_{\rm PQ}}\sim\mathcal{O}(10^3)$
avoiding  symmetry restoration after inflation.

\subsubsection{Benchmark II}
\label{sec:reduced_backreaction_benchmark}

For the first benchmark, we took values of the parameters such that the evolution of the PQ vacuum changes the
PQ-sector contribution to the total vacuum energy by approximately
$\mathcal{O}(10\%)$ of the inflationary energy density. Although the PQ-sector energy is subdominant in this case, it is interesting to investigate a benchmark that is even more conservative.
We therefore consider a second benchmark with smaller magnitudes of $\lambda(\Lambda)$ and $c_6$:
\begin{equation}
    \lambda(\Lambda)
    =
    -7\times10^{-7},
    \qquad
    y(\Lambda)
    =
    0.05,
    \qquad
    c_6
    =
    1.2\times10^{-5}.
    \label{eq:para_21}
\end{equation}
\begin{figure}[t]
    \centering
    \includegraphics[width=0.425\linewidth]{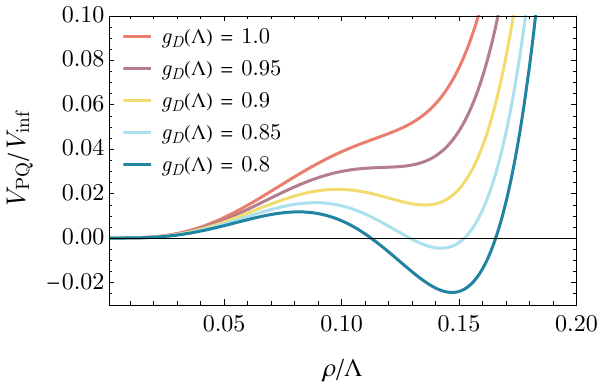}
    \includegraphics[width=0.4\linewidth]{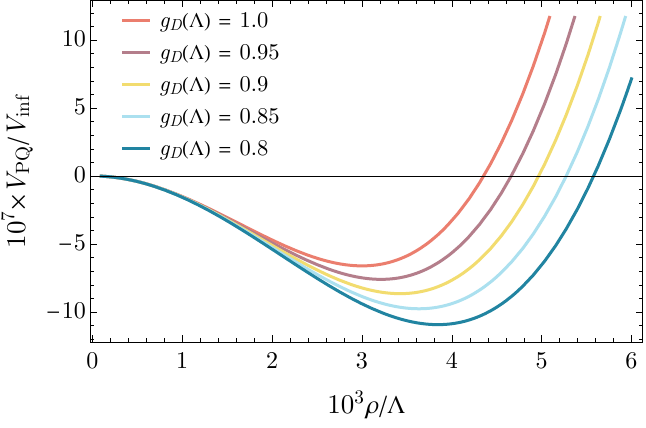}
    \caption{
    Left panel: The RG-improved PQ potential in the vicinity of the initial
    large field vacuum $\rho_{\rm min}$ for several boundary values of the hidden gauge
    coupling $g_D(\Lambda)$, for Benchmark II defined in Eq.~\eqref{eq:para_21}. The vertical axis is normalized to the
    inflationary energy density
    $V_{\rm inf}=3M_{\rm pl}^2H_{\rm inf}^2$. As
    $g_D(\Lambda)$ increases, $\rho_{\rm min}$ moves toward
    smaller values, becomes shallower, and eventually disappears. Right panel: The RG-improved potential in the vicinity of $\widetilde{\rho}_{\rm min}$ for the same set of values of $g_D(\Lambda)$.
    }
    \label{fig:V_gD_illustration_2}
\end{figure}
\begin{figure}[ht]
    \centering
    \includegraphics[width=0.99\linewidth]{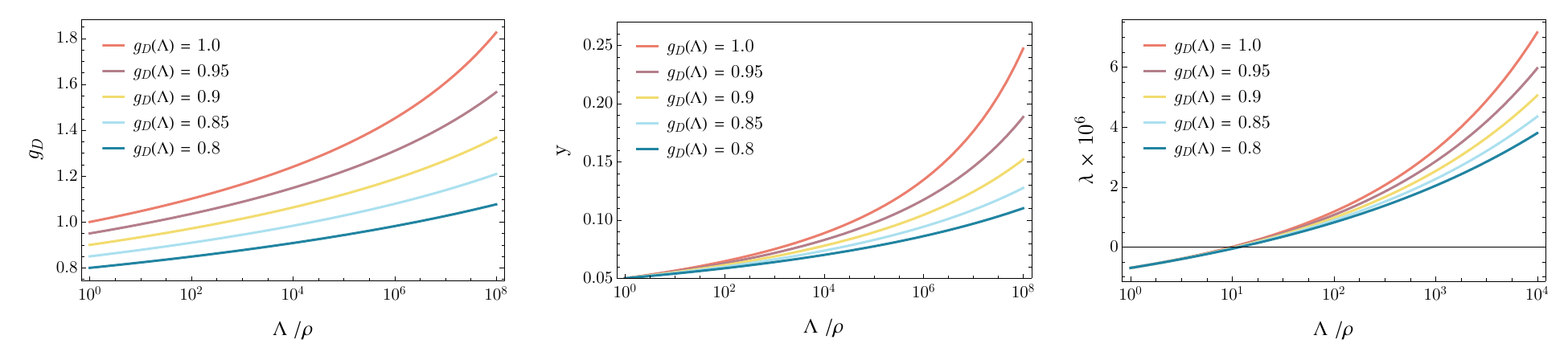}
    \caption{
    The running couplings $g_D(\rho)$, $y(\rho)$, and $\lambda(\rho)$
    as functions of $\Lambda/\rho$ for Benchmark II, defined in
    Eq.~\eqref{eq:para_21}. Moving to the right corresponds to running
    from the cutoff scale toward smaller field values. The different
    curves correspond to different boundary values of
    $g_D(\Lambda)$.
    }
    \label{fig:Couplings_illustration_2}
\end{figure}
As shown in Fig.~\ref{fig:V_gD_illustration_2} and \ref{fig:Couplings_illustration_2}, the PQ vacuum transition changes the potential energy by only a few percent of $V_{\rm inf}$ due to the choice of smaller $\lambda, y$ and $c_6$ values. As a result, the smaller value of the quartic coupling at the intermediate field range drives $\widetilde{\rho}_{\rm min}$ to a larger value. 
This benchmark therefore illustrates that the removal of the large field minimum, controlled by the dynamics of the inflaton, need not require
an order one transfer of energy between the inflaton and PQ sectors.

\subsection{Inflaton-Dependent Yukawa Coupling}
\label{sec:inflaton_dependent_yukawa_coupling}

We now turn to our second model of an inflaton-controlled transition from $\rho_{\rm min}$ to $\widetilde{\rho}_{\rm min}$. The model avoids a coupling of the inflaton to a hidden gauge sector, instead introducing an inflaton-dependent Yukawa operator for the fermions. The fermionic sector of the Lagrangian is given by
\begin{equation}\label{yukawamodel}
   \mathcal{L}_{\rm Yuk} =  \brr{y+\kappa\dfrac{\phi}{\Lambda}} \brr{\Phi\, \bar{\psi}_L \psi_R + \text{h.c.}} \ .
\end{equation}
We do not assume that the  fermion carries the hidden group charge. Therefore the RGE equations in this scenario can be equivalently obtained by setting $N = 1$ and $g_D = 0$ in Eq.~\eqref{eq:beta_lambda} and \eqref{eq:beta_y}. 

This inflaton-controlled dimension-5 operator can be obtained by integrating out a heavy scalar or fermion. The Yukawa coupling $y$ at the boundary UV scale $\Lambda$ is therefore modulated by the inflaton rolling, specifically
\begin{equation}\label{yukawarolling}
   y(\Lambda,\phi) = y_0 + \kappa\dfrac{\phi}{\Lambda} \ .  
\end{equation}
Here $y_0$ is the Yukawa coupling at $\phi = 0$. As $\kappa \,\phi$  increases, the size of $y(\Lambda)$  also increases, which helps to raise the large field PQ vacuum.

As in the model with the inflaton dependent gauge coupling, we will require that the PQ sector contributes an energy that is subdominant to the energy in the inflationary sector: $|V_{\rm PQ}(\rho_{\rm min})|/V_{\rm inf} < \mathcal{O}(10\%)$. 

At the location of the intermediate vacuum $\widetilde{\rho}_{\rm min}$, the requirement $\widetilde{\rho}_{\rm min}/v_{\rm PQ} < 10^4$ right before the end of inflation is more easily satisfied in this scenario than in the model in Section~\ref{sec:inflaton_dependent_gauge_coupling}. The reason is as follows. From Eq.~\eqref{yukawarolling}, it is clear that as the inflaton rolls, the Yukawa coupling is enhanced: $y(\Lambda)$ at a later stage of inflation is larger than that in an earlier era. Remembering from Eq.~\eqref{eq:beta_gD} that the $\beta$ function of the quartic coupling $\lambda$ is sensitive to $y^4$, $\lambda(\widetilde{\rho}_{\rm min})$ is thus correspondingly pushed to larger values, which means that  $\widetilde{\rho}_{\rm min}$ can be smaller. This should be contrasted with the dilaton-like operator discussed previously, for which it was the increasing hidden gauge coupling that enhanced the Yukawa coupling, instead of the direct rolling of the inflaton field.

 In the following, we will display two benchmark cases. In the first, we will choose larger magnitude of $\lambda$ and $c_6$ so that $|\Delta V_{\rm PQ}(\rho_{\rm min})|/V_{\rm inf} \simeq \mathcal{O}(10\%)$. 
While for the second benchmark, the smaller couplings are adopted and $|\Delta V_{\rm PQ}(\rho_{\rm min})|/V_{\rm inf}$ is at the percent level.

\begin{figure}[ht]
    \centering
    \includegraphics[width=0.86\linewidth]{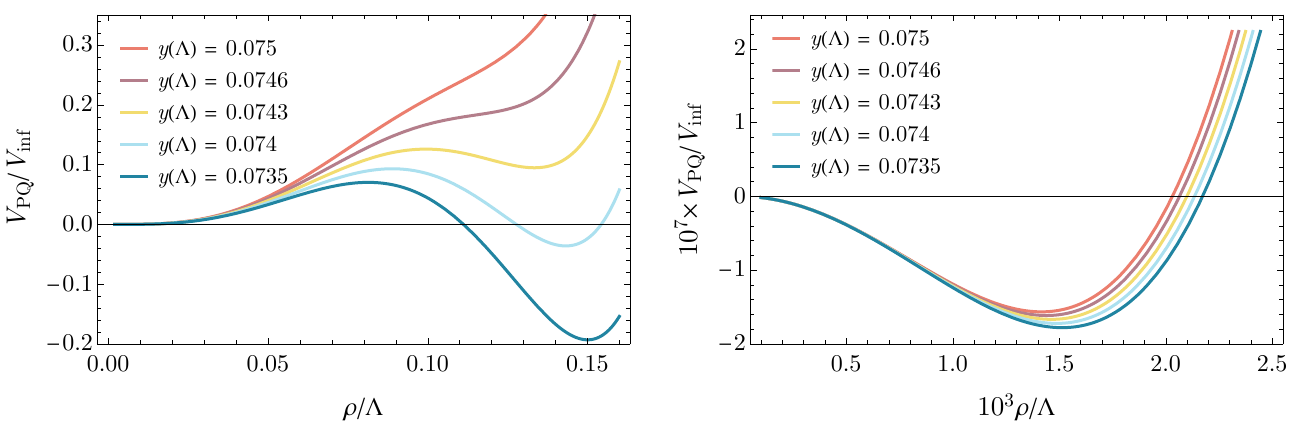}\\
    \includegraphics[width=0.9\linewidth]{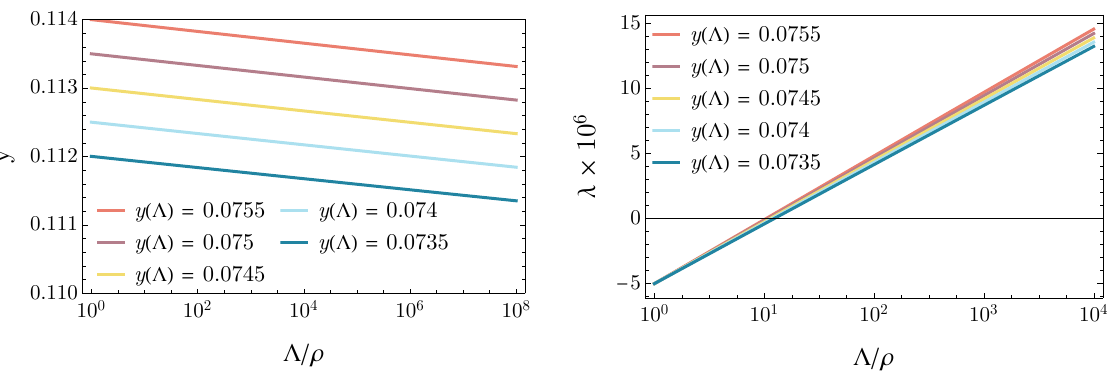}
    \caption{ The PQ field profiles and the running coupling under the parameter choice in 
    Eq.~\eqref{eq:B_para1}. Curves of different colors correspond to different Yukawa coupling choices at the boundary scale $\Lambda$.}
    \label{fig:B_1}
\end{figure}

\subsubsection{Benchmark I}

We first choose
\begin{equation}\label{eq:B_para1}
    \lambda(\Lambda) = -5 \times 10^{-6}, \qquad c_6 = 7.64 \times 10^{-5} \ .
\end{equation}
The Yukawa coupling at the boundary scale is varied over the range
$y(\Lambda)=0.0735$--$0.075$, as shown in Fig.~\ref{fig:B_1}.
The upper left and right panels illustrate the PQ potential in the vicinity of $\rho_{\rm min}$ and $\widetilde{\rho}_{\rm min}$, respectively, for five different $y(\Lambda)$ values.
The overall PQ field evolution is similar to that in Section~\ref{sec:inflaton_dependent_gauge_coupling}.
The lower left and right panels show the
corresponding running of $y$ and $\lambda$, respectively, as a function of $\Lambda/\rho$.
It should be noted that the Yukawa coupling $y$ decreases as the running scale drops due to the positive $\beta_y$ function, which is different from that in Fig.~\ref{fig:Couplings_illustration_2}.  

Although the running of the Yukawa coupling itself is mild, the small
variation of its boundary value has a sizable effect on the RG evolution
of the quartic coupling.  In particular, increasing $y(\Lambda)$ leads
to a faster increase of $\lambda$ toward smaller renormalization scales,
which progressively lifts the large-field region of the PQ potential.
For the smaller values of $y(\Lambda)$, the potential contains a
large-field minimum at $\rho/\Lambda\sim {\cal O}(0.1)$.
As $y(\Lambda)$ is increased, this minimum is lifted and becomes
progressively shallower, before disappearing once a critical value of
$y(\Lambda)$ is crossed.  In contrast, the small-field minimum
$\widetilde{\rho}_{\rm min}\sim {\cal O}(10^{-3})\Lambda$ persists
throughout the scan and only shifts mildly.
The transition between the two vacuum structures therefore occurs
within a rather narrow interval of $y(\Lambda)$.

We can set $y(\Lambda) = 0.0735$ at the inflationary CMB window stage and enhance $y(\Lambda)$ to around 0.075 at the end of inflation. Therefore, the inflaton rolling needs to satisfy
\begin{equation}
    \kappa \dfrac{\Delta \phi}{\Lambda} = 0.0015 \ .
\end{equation}

\subsubsection{Benchmark II}

\begin{figure}[th]
    \centering
    \includegraphics[width=0.86\linewidth]{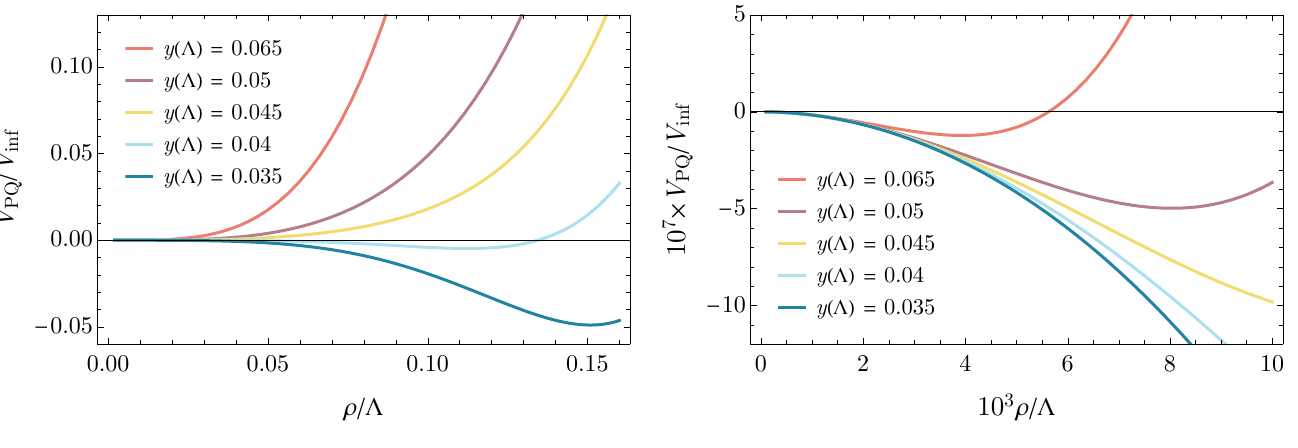}\\
    \includegraphics[width=0.9\linewidth]{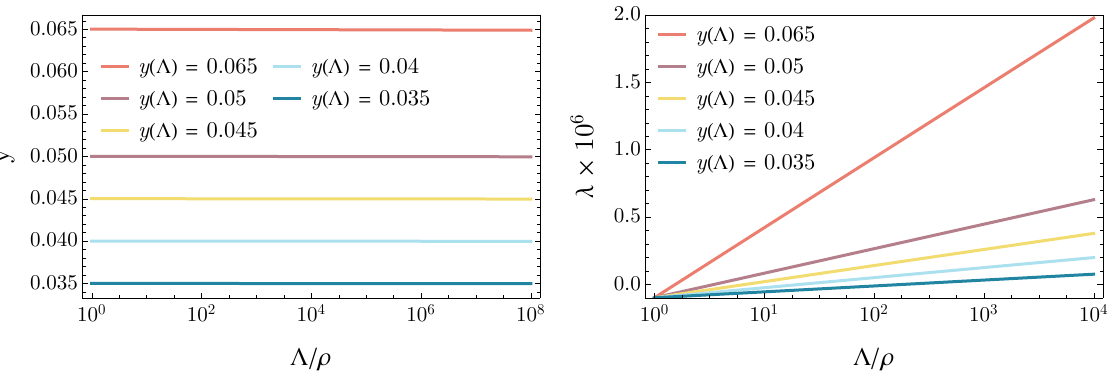}
    \caption{The PQ field profiles and the running coupling under the parameter choice in 
    Eq.~\eqref{eq:B_para2}. Different color of curves correspond to different Yukawa coupling choices at the boundary scale $\Lambda$.   }
    \label{fig:B_2}
\end{figure}

 We next consider a different choice of  parameters. Our second benchmark consists of the set of parameters,
\begin{equation}\label{eq:B_para2}
    \lambda(\Lambda) = -  10^{-7}, \qquad c_6 = 3.05 \times 10^{-6} \ .
\end{equation}
Compared to Benchmark I, both $|\lambda(\Lambda)|$ and $c_6$ are
reduced by an order of magnitude. The Yukawa coupling $y(\Lambda)$ is varied from 0.035 to 0.065 in a broader range, as shown in Fig.~\ref{fig:B_2}.
As $y(\Lambda)$ crosses a critical value slightly larger than 0.04, the large field vacuum $\rho_{\rm min}$ disappears. 
While the intermediate vacuum $\widetilde{\rho}_{\rm min}$ remains at the order of $\mathcal{O}(10^{-3}-10^{-2})M_{\rm pl}$. 
To satisfy the requirement in
Eq.~\eqref{eq:intermediate_final_benchmark_ratio},  $y(\Lambda)$ at the end of inflation is taken to be around 0.065 as a representative choice, so that $\widetilde{\rho}_{\rm min}$ can be smaller than $10^{4} f_a$. While larger values of $y(\Lambda)$ are
also allowed, which can further reduce $\widetilde{\rho}_{\rm min}$. If we choose $y(\Lambda) = 0.04$ at the inflationary CMB window, the inflaton rolling range satisfies 
\begin{equation}
    \kappa \dfrac{\Delta \phi}{\Lambda} \simeq 0.02 \ .
\end{equation}
%

Compared to the previous dilaton-like operator in Sec.~\ref{sec:inflaton_dependent_gauge_coupling}, this scenario requires less tuning, provided that the variation of $y(\Lambda)$ is sufficient to cross the threshold at which the large-field vacuum disappears.

\subsection{Back-reaction on the Inflaton Potential}\label{backreactioninflaton}

\begin{figure}[h]
    \centering
    \includegraphics[width=0.75\linewidth]{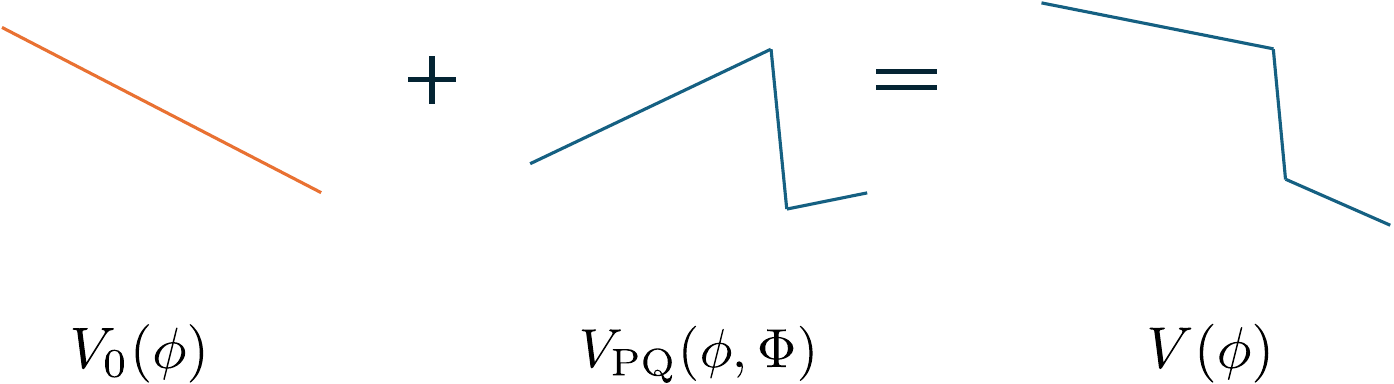}
    \caption{The sketch of the vacuum potential composed of $V_{0}(\phi)$ and $V_{\rm PQ}(\phi,\Phi)$.}
    \label{fig:V_illu}
\end{figure}

For the benchmark parameters considered above, the variation of the PQ vacuum energy satisfies
$\Delta V_{\rm PQ}/V_{\rm inf}\lesssim \mathcal{O}(10\%)$. The effective potential along the inflationary trajectory can be decomposed as
\begin{equation}
    V(\phi) = V_{0}(\phi) + V_{\rm PQ}(\phi,\Phi) \ .
\end{equation}
The first term $V_{0}(\phi)$ denotes the contribution from the inflaton sector, which is independent of the PQ vacuum configuration. 
The second term represents the PQ vacuum energy evaluated at the instantaneous minimum $\Phi_{\rm min}(\phi)$. Its dependence on $\phi$ arises indirectly through the dimension-five operator that controls the evolution of the PQ potential. A schematic illustration of these two contributions and their sum is shown in Fig.~\ref{fig:V_illu}.

For the inflaton to continue rolling in the desired direction, the slope of $V_{0}(\phi)$ must be sufficiently large to overcome the opposing variation of the PQ vacuum energy. More precisely, the total slope
\begin{equation}
V'(\phi)
=
V_{0}'(\phi)
+
\frac{d}{d\phi}
V_{\rm PQ}\bigl(\phi,\Phi_{\rm min}(\phi)\bigr)
\end{equation}
must retain the appropriate sign throughout the transition. 
It is important to emphasize that the individual contribution from the PQ sector to the inflaton potential and its derivatives need not be small by itself. Rather, the inflationary observables are determined by the total effective potential in Eq.~(4.22). In particular, the contribution from the PQ vacuum energy may partially cancel the corresponding slope and curvature of $V_0(\phi)$, leading to suppressed derivatives of the total potential. We therefore only require that, over the field range associated with the observable CMB modes, the combined potential $V(\phi)$ yields an inflationary evolution consistent with the CMB constraints.

As the large-field minimum at $\rho_{\rm min}$ disappears, the PQ field evolves toward the intermediate minimum $\widetilde{\rho}_{\rm min}$. This change lowers the total vacuum energy by at most $\mathcal{O}(10\%)$ for the benchmarks in the last section. 
Such a localized feature is phenomenologically acceptable provided that the transition occurs sufficiently far from the field range corresponding to the observable CMB window, so that it does not leave any imprints on the measured scalar power spectrum.

After the transition from $\rho_{\rm min}$ to $\widetilde{\rho}_{\rm min}$, the PQ sector contributes only a subdominant fraction of the total vacuum energy and its slope. 
Consequently, the subsequent inflaton evolution is governed predominantly by $V_{0}(\phi)$. 
Although the position of $\widetilde{\rho}_{\rm min}$ continues to vary with $\phi$, its motion induces negligible backreaction on the inflationary trajectory.
The partial cancellation between the inflaton and PQ sectors that operates during the CMB-relevant stage is almost removed. As a result, the magnitude of the inflaton slope and the slow-roll parameter $\epsilon_V$ increases, leading to a somewhat larger inflaton velocity and therefore a suppressed scalar curvature power spectrum on the smaller scales that exit the horizon subsequently.

Overall, a natural cosmological history is that the inflaton evolves only slowly over the observable CMB window, so that the corresponding variation of the PQ vacuum configuration remains small during this stage. The partial cancellation between $V_0(\phi)$ and $V_{\rm PQ}\bigl(\phi,\Phi_{\min}\bigr)$ can then be arranged such that the total effective potential satisfies the CMB constraints on the slow-roll parameters. After the observable CMB modes have exited the horizon, the shape of $V_0(\phi)$ may evolve such that this cancellation becomes progressively weaker. The resulting increase in the total inflaton slope leads to a larger $|\dot{\phi}|$, allowing the inflaton to scan the PQ-sector parameters more rapidly and to reach the critical value at which the large-field minimum disappears within a relatively short interval of e-folds. In this way, the PQ sector can remain nearly unchanged throughout the CMB-relevant stage while undergoing a much more rapid evolution only at a later epoch, while still allowing the total inflaton excursion to remain sub-Planckian. The transition itself is temporarily a genuinely two-field process, whereas before and after the transition the evolution can again be described by an effective single-field trajectory, provided that the PQ field adiabatically follows the corresponding minimum.

\section{Transition Dynamics and Open Issues}
\label{sec:transition_dynamics}

We have demonstrated in the last section that
the initial vacuum $\rho_{\rm min}$  can be removed through a
controlled evolution of the RG-improved PQ potential, by relying on the dynamics of the inflaton $\phi$. In this Section, we discuss the actual dynamics of the saxion field as it transitions from $\rho_{\rm min}$ to $\widetilde\rho_{\rm min}$, and explore the conditions under which a classical transition, as opposed to a phase transition, occurs.

Depending on the shape of the potential and the rate at which it evolves, the PQ field may leave the initial vacuum through classical rolling, quantum
tunneling, or a combination of the two. For the mechanism proposed in our work, the preferred history is a classical
transition in which the high-field minimum $\rho_{\rm min}$ and the intervening barrier to the stepping stone $\widetilde{\rho}_{\rm min}$ disappear before vacuum decay becomes appreciable. The radial field then rolls toward the intermediate minimum while the universe is still
inflating. Hubble friction and the remaining inflationary expansion can damp the homogeneous motion and dilute the fluctuations generated during the transition.  A first order transition is not necessarily incompatible with the
mechanism, but it introduces additional questions involving bubble nucleation, bubble collisions, nonthermal PQ restoration, and defect
formation.

From the left panel of Fig.~\ref{fig:V_gD_illustration}, it is evident that the PQ field potential barrier height keeps decreasing as the rolling of inflaton raises the hidden gauge coupling.
Before the barrier disappears completely, the initial large-field vacuum $\rho_{\rm min}$ 
may become metastable. The tunneling rate per unit physical volume can be
written parametrically as
\begin{equation}
    \frac{\Gamma_{\rm tun}}{\mathcal{V}}
    \simeq
    A\, e^{-B},
    \label{eq:tunneling_rate}
\end{equation}
where $B$ is the Euclidean bounce action and $A$ is a fluctuation prefactor.  We can first make a naive qualitative estimation. 
Denoting the position of the barrier top as $\rho_{\rm t}$, we define
\begin{equation}
    \Delta \rho = \rho_{\rm min} - \rho_{\rm t}, \qquad
    \Delta V_{\rm bar} = V_{\rm PQ}(\rho_{\rm t}) - V_{\rm PQ}(\rho_{\rm min}) \ .
\end{equation}
It should be noted that $\rho_{\rm min}$ is slowly varying with time. The action can be roughly estimated as
\begin{equation}
  B \sim C \dfrac{\brr{\Delta \rho}^4}{\Delta V_{\rm bar}}   \ ,
\end{equation}
in which the coefficient $C$ is at the order of $\mathcal{O}(1\sim 100)$. For the benchmark values of $\Delta \rho \sim \mathcal{O}(0.01) M_{\rm pl}$ and $\Delta V_{\rm bar}  \simeq \mathcal{O}(0.01) V_{\rm inf}$, the action $B$ is approximately at the order of $\mathcal{O}(10^4)$ to $\mathcal{O}(10^6)$. Therefore the quantum tunneling rate is highly suppressed.

Unless $\Delta \rho$ is extremely small, the tunneling might take place just  before the large field vacuum position completely disappears, which is similar to the double infield inflation~\cite{Adams:1990ds}. However, this requires  fine tuning  both the potential parameters and the inflaton rolling velocity. The inflaton velocity $\dot{\phi}$ should be decelerated to a very tiny value to guarantee the relatively long duration at this point so that the bubbles can be nucleated and expand. 
If the bubbles can successfully merge and the phase transition can be completed,  gravitational waves from  bubble collision would constitute a viable signal potentially detectable by future  observatories.
The GW spectrum is expected to display the oscillatory pattern~\cite{An:2020fff,An:2022cce,Zou:2026wzi,Hu:2025xdt,Liang:2026wwz}. The large curvature perturbation from the FOPT can also induce second order GW signal~\cite{An:2023jxf}. 
However, it is also possible that large bubbles fail to merge but are quickly frozen upon exiting the horizon.  In this case, the phase transition would be incomplete~\cite{Barir:2022kzo}  and the universe would remain in the false vacuum. The background PQ field would still continue with its classical rolling after the large field vacuum disappears.

When the PQ field evolves through classical rolling rather than tunneling, parametric resonance may nevertheless generate an observable stochastic GW background. In particular, resonantly amplified PQ fluctuations can develop sizable anisotropic stress, which directly sources tensor perturbations. The same dynamics may also enhance scalar fluctuations and produce a localized peak in the curvature perturbation power spectrum. After inflation, when these enhanced modes re-enter the horizon, they can generate an additional scalar induced GW signal at second order.

In our scenario, the approximately quartic PQ potential can yield efficient parametric resonance. Owing to the approximate conformal invariance of the quartic system, cosmic expansion does not rapidly shift the amplified modes out of the resonance bands in comoving momentum space. Nevertheless, the rapid cosmic expansion can largely dilute the physical amplitudes of the PQ fluctuations, while their resonant growth requires a finite duration to accumulate. 
Together with the requirement that the PQ symmetry not be nonthermally restored, these effects are expected to lead the resulting curvature perturbations small. 
The associated scalar induced GW signal should therefore be highly suppressed, although the amplified PQ fluctuations may still directly source gravitational waves through their anisotropic stress.

GW production during preheating has been extensively studied in the literature~\cite{Garcia:2024zir,del-Corral:2025fzz}. To the best of our knowledge, however, the corresponding dynamics associated with parametric amplification of a scalar field during inflation has not yet been investigated in detail. A reliable determination of the resulting GW amplitude and spectrum, as well as the associated curvature perturbations, requires a dedicated nonlinear lattice simulation that consistently accounts for the coupled evolution of the inflaton, the PQ field, and metric perturbations. We leave such an analysis for future work.

\section{Conclusion and Outlook}\label{sec:conclusion}

In this work, we have proposed a new realization of the varying-PQ-scale solution to the axion isocurvature problem. A generic challenge for this class of scenarios is the possible nonthermal restoration of the PQ symmetry after inflation. If the PQ field rolls directly from a large field value $\rho_{\rm min}$ during inflation to the present-day vacuum $v_{\rm PQ}$, the resulting large-amplitude oscillations can efficiently amplify radial and angular fluctuations through parametric or tachyonic resonance. Once these fluctuations become sufficiently large, the PQ symmetry may be nonthermally restored, leading to the regeneration of axion strings and domain walls. To alleviate this problem, we divide the evolution of the PQ field into two stages by introducing an intermediate minimum $\widetilde{\rho}_{\rm min}$.

This structure can be realized by taking the PQ quartic coupling to be negative at the ultraviolet boundary scale $\Lambda$. Renormalization-group evolution can then drive the quartic coupling positive at lower scales. Within an appropriate region of parameter space, the RG-improved PQ potential consequently develops two distinct local minima: a large-field minimum at $\rho_{\rm min}$ and an intermediate-field minimum at $\widetilde{\rho}_{\rm min}$. The transition between these minima substantially reduces the field excursion during the subsequent evolution toward $v_{\rm PQ}$ and thereby helps avoid nonthermal PQ symmetry restoration. 

To dynamically modify the profile of the PQ potential, we introduce dimension-five interactions involving the inflaton. In the first realization, the PQ-charged fermions are assumed to carry charges under an additional strongly interacting hidden gauge group, and the inflaton couples to the corresponding gauge kinetic term. 
As the inflaton rolls, it changes the hidden gauge coupling, which in turn modifies the RG evolution of the Yukawa coupling. For suitable boundary conditions, the resulting increase in the Yukawa coupling removes the large-field minimum at $\rho_{\rm min}$. This realization, however, requires the UV Yukawa coupling to lie within a relatively narrow range and therefore involves a certain degree of parameter tuning.

Alternatively, we consider a dimension-five interaction through which the inflaton directly modifies the Yukawa coupling. In this case, the inflaton directly scans the Yukawa coupling, allowing a broader range of UV boundary values and reducing the required tuning. Both realizations provide a dynamical connection between the inflaton evolution and the disappearance of the large-field PQ minimum.

The rolling PQ field may undergo parametric resonance and source a stochastic gravitational-wave background through the anisotropic stress of the amplified fluctuations. The same dynamics may generate a localized enhancement in the curvature-perturbation power spectrum and, consequently, a scalar-induced gravitational-wave signal when the enhanced modes later re-enter the horizon. In the approximately quartic PQ potential considered here, cosmic expansion does not rapidly shift the relevant comoving modes out of the resonance bands. Nevertheless, it continuously dilutes their physical amplitudes, while resonant growth requires a finite amount of time to accumulate. Together with the requirement that the PQ symmetry not be nonthermally restored, these effects are expected to keep the induced curvature perturbations small and hence strongly suppress the scalar-induced gravitational-wave component. A reliable determination of the directly sourced gravitational-wave spectrum requires a dedicated nonlinear lattice simulation, which we leave for future work.

Compared with previous realizations of the varying-PQ-scale mechanism, our construction does not require an extremely small quartic coupling of order $10^{-14}$. Viable examples instead admit a quartic coupling of order $10^{-7}$--$10^{-6}$ together with a percent-level Yukawa coupling. Our results therefore open a new possibility for realizing the varying-PQ-scale solution with a non-flat PQ potential and comparatively less hierarchical model parameters.

An interesting extension of our framework is to employ the same dimension-five interactions as a dynamical mechanism for terminating inflation, in analogy to the hybrid inflation~\cite{Linde:1991km,Linde:1993cn,Copeland:1994vg,Cortes:2009ej,Ashoorioon:2015hya}. In such a realization, the PQ sector may provide the dominant inflationary vacuum energy, while a slowly rolling spectator field scans the RG-improved PQ potential through its dimension-five coupling. Once the spectator reaches a critical value, the large-field PQ minimum disappears, triggering the rapid classical rolling of the PQ field and thereby ending inflation. This setup resembles a spectator-controlled or modulated end of inflation. Its viability requires a dedicated analysis of the curvature perturbations generated by fluctuations of the spectator field, the duration of the PQ rolling stage, and the subsequent reheating dynamics. We leave a detailed investigation of this possibility for future work.

\acknowledgments

This work was  performed in part at the Aspen Center for Physics, which is supported by National Science Foundation grant PHY-2210452. We would like to thank the organizers of the Center for Theoretical Underground Physics and Related Areas (CETUP*) for their hospitality and financial support during the course of this work.
K.-F.Lyu acknowledges Jiji Fan and Matthew Reece for useful discussions. The research activities of K.S. are supported in part by the U.S. National Science Foundation under Award No. PHY-2412671.

\appendix
\section{Evolution of Saxion Field Fluctuations}

We consider a time-dependent PQ potential,
\begin{equation}
V(\rho,t)
=
V\bigl(\rho,\phi(t),H(t)\bigr),
\end{equation}
where the explicit time dependence may arise from the inflaton background and Hubble-induced terms. The full nonlinear equation of
motion in a spatially flat FLRW background is
\begin{equation}
\ddot{\rho}
+3H(t)\dot{\rho}
-\frac{\nabla^2\rho}{a^2(t)}
+V_{,\rho}(\rho,t)
=0.
\end{equation}
We decompose the field into a homogeneous background and an inhomogeneous
perturbation,
\begin{equation}
\rho(t,\mathbf{x})
=
\bar{\rho}(t)+\delta\rho(t,\mathbf{x}),
\qquad
\langle\delta\rho(t,\mathbf{x})\rangle=0.
\end{equation}
The background $\bar{\rho}(t)$ is the actual homogeneous solution and may not be identified with the instantaneous minimum of the potential.
Neglecting the backreaction of the fluctuations, it satisfies
\begin{equation}
\ddot{\bar{\rho}}
+3H(t)\dot{\bar{\rho}}
+V_{,\rho}\bigl(\bar{\rho}(t),t\bigr)
=0.
\end{equation}
Expanding the full equation to linear order in $\delta\rho$ gives
\begin{equation}
\delta\ddot{\rho}
+3H(t)\delta\dot{\rho}
-\frac{\nabla^2\delta\rho}{a^2(t)}
+m_{\rm eff}^2(t)\delta\rho
=0,
\end{equation}
where
\begin{equation}
m_{\rm eff}^2(t)
\equiv
V_{,\rho\rho}\bigl(\bar{\rho}(t),t\bigr).
\end{equation}
Thus, the effective mass must be evaluated along the actual background
trajectory rather than at the instantaneous minimum, unless the background
adiabatically tracks that minimum.
Using the Fourier decomposition
\begin{equation}
\delta\rho(t,\mathbf{x})
=
\int\frac{d^3k}{(2\pi)^3}
\delta\rho_k(t)e^{i\mathbf{k}\cdot\mathbf{x}},
\end{equation}
each mode obeys
\begin{equation}
\delta\ddot{\rho}_k
+3H(t)\delta\dot{\rho}_k
+
\left[
\frac{k^2}{a^2(t)}
+m_{\rm eff}^2(t)
\right]
\delta\rho_k
=0.
\end{equation}

During inflation, if $H$ and $m_{\rm eff}$ are approximately constant, the
super-Hubble solution can be written as
\begin{equation}
\delta\rho_k
=
C_{k,+}a^{-3/2+\nu}
+
C_{k,-}a^{-3/2-\nu},
\qquad
\nu
=
\sqrt{
\frac{9}{4}
-\frac{m_{\rm eff}^2}{H^2}
}.
\end{equation}
For $m_{\rm eff}<3H/2$, the dominant mode therefore evolves as
\begin{equation}
\delta\rho_k
\propto
a^{-\Delta},
\qquad
\Delta
\equiv
\frac{3}{2}-\nu.
\end{equation}
In the light-field limit,
$m_{\rm eff}^2\ll H^2$,
one has
\begin{equation}
\Delta
\simeq
\frac{m_{\rm eff}^2}{3H^2},
\end{equation}
and hence
\begin{equation}
\delta\rho_k(t_{\rm end})
\simeq
\delta\rho_k(t_*)
\exp\left[
-\int_{N_*}^{N_{\rm end}}
dN\,
\frac{m_{\rm eff}^2(N)}{3H^2(N)}
\right].
\end{equation}
For a Bunch--Davies initial state,
\begin{equation}
\mathcal{P}_{\delta\rho}^{1/2}(k,t_*)
\simeq
\frac{H_*}{2\pi},
\end{equation}
so that
\begin{equation}
\mathcal{P}_{\delta\rho}^{1/2}(k,t_{\rm end})
\simeq
\frac{H_*}{2\pi}
\exp\left[
-\int_{N_*}^{N_{\rm end}}
dN\,
\frac{m_{\rm eff}^2(N)}{3H^2(N)}
\right],
\end{equation}
provided the light-field and adiabatic approximations remain valid. While for certain range of modes, the fluctuation amplitude can be enhanced by the parametric resonance amplification during the PQ field rolling from $\rho_{\rm min}$ to $\tilde{\rho}_{\rm min}$.

After inflation, both $H(t)$ and $m_{\rm eff}(t)$ evolves over time. The
background and fluctuation equations must therefore be solved together,
\begin{align}
\ddot{\bar{\rho}}
+3H(t)\dot{\bar{\rho}}
+V_{,\rho}(\bar{\rho},t)
&=0,
\\
\delta\ddot{\rho}_k
+3H(t)\delta\dot{\rho}_k
+
\left[
\frac{k^2}{a^2(t)}
+V_{,\rho\rho}(\bar{\rho},t)
\right]
\delta\rho_k
&=0.
\end{align}
The fluctuation $\delta\rho_k$ is therefore defined with respect to the
time-dependent homogeneous solution $\bar{\rho}(t)$, not with respect to a
fixed vacuum value such as $f_a$.

For a mode that remains outside the Hubble radius,
$k/a\ll H$,
and satisfies
$m_{\rm eff}^2\ll H^2$,
the overdamped approximation gives
\begin{equation}
3H\delta\dot{\rho}_k
+m_{\rm eff}^2\delta\rho_k
\simeq0,
\end{equation}
with solution
\begin{equation}
\delta\rho_k(t)
\simeq
\delta\rho_k(t_{\rm end})
\exp\left[
-\int_{N_{\rm end}}^{N}
dN'\,
\frac{m_{\rm eff}^2(N')}{3H^2(N')}
\right].
\end{equation}
If instead $m_{\rm eff}\gtrsim H$, the fluctuation oscillates and its envelope
typically decreases approximately as
\begin{equation}\label{eq:evo_large_meff}
|\delta\rho_k|
\propto
a^{-3/2}m_{\rm eff}^{-1/2},
\end{equation}
in the adiabatic regime. 

Assuming the universe is in the matter-dominated era after inflation, the Hubble evolves like $H(t)\sim 2/(3t)\sim a(t)^{-3/2}$. If the Hubble induced mass dominates, 
\begin{equation}
  \dfrac{m_{\rm eff}^2(N)}{3 H^2(N)} = 2\, \xi  \ . 
\end{equation}
The dilution factor is 
\begin{equation}
    \exp\left[
-\int^{N}_{N_{\rm end}}
dN\,
\frac{m_{\rm eff}^2(N)}{3H^2(N)}
\right] \sim \brr{\dfrac{a_{\rm end}}{a(N)}}^{2\xi} \ .
\end{equation}
Once the Hubble induced mass is sub-dominant, $m_{\rm eff} \geq H$, the fluctuation would dilute as in Eq.~\eqref{eq:evo_large_meff}.

The horizon re-entry time for a given mode is defined by $k/a(t_{\rm re})=
H(t_{\rm re})$.
Its fluctuation amplitude at re-entry is
\begin{equation}
\Delta_\rho(k,t_{\rm re})
\equiv
\mathcal{P}_{\delta\rho}^{1/2}(k,t_{\rm re})
=
\sqrt{\frac{k^3}{2\pi^2}}
\left|
\delta\rho_k(t_{\rm re})
\right|.
\end{equation}
Horizon re-entry itself does not amplify the mode. In the absence of
tachyonic or parametric instabilities, a positive effective mass either leaves
a sufficiently light super-Hubble mode approximately constant or further
suppresses it.
A rough single-mode criterion for possible PQ restoration after the background
has approached the present vacuum is
\begin{equation}
\Delta_\rho(k,t_{\rm re})
\gtrsim v_{\rm PQ}.
\end{equation}

\bibliographystyle{utphys}
\bibliography{ref}

\end{document}